# All-glass 100 mm Diameter Visible Metalens for Imaging the Cosmos

**Authors:** Joon-Suh Park[1]*†, Soon Wei Daniel Lim[1]†, Arman Amirzhan[1], Hyukmo Kang[2], Karlene Karrfalt[2,3], Daewook Kim[2], Joel Leger[3], Augustine M. Urbas[3], Marcus Ossiander[1,4], Zhaoyi Li[1], and Federico Capasso[1]*

[1]John A. Paulson School of Engineering and Applied Sciences, Harvard University; Cambridge, Massachusetts, 02138, United States.

[2]Wyant College of Optical Sciences, The University of Arizona; Tucson, Arizona, 85721, United States.

[3]Air Force Research Laboratory, Wright-Patterson Air Force Base; Dayton, Ohio 45433, United States.

[4]Institute of Experimental Physics, Graz University of Technology, 8010 Graz, Austria.

†These authors contributed equally to this work.

*Corresponding author. Email: parkj@g.harvard.edu, capasso@seas.harvard.edu

†: These authors equally contributed to this work.

**Abstract**:

Metasurfaces, optics made from subwavelength-scale nanostructures, have been limited to millimeter-sizes by the scaling challenge of producing vast numbers of precisely engineered elements over a large area. In this study, we demonstrate an all-glass 100 mm diameter metasurface lens (metalens) comprising 18.7 billion nanostructures that operates in the visible spectrum with a fast f-number (*f*/1.5, NA=0.32) using deep-ultraviolet (DUV) projection lithography. Our work overcomes the exposure area constraints of lithography tools and demonstrates that large metasurfaces are commercially feasible. Additionally, we investigate the impact of various fabrication errors on the imaging quality of the metalens, several of which are unique to such large area metasurfaces. We demonstrate direct astronomical imaging of the Sun, the Moon, and emission nebulae at visible wavelengths and validate the robustness of such metasurfaces under extreme environmental thermal swings for space applications.

## Introduction:

Large objective apertures are necessary for optical systems which collect weak or rapidly evolving signals, such as in astronomical imaging and remote airborne surveillance, as well as in high-power laser applications to decrease the incident power density. Due to the optomechanical mounting limitations and the volumetric-and-weight scaling of refractive objective lenses with the diameter, large aperture optics must often be reflective instead of transmissive. Dielectric metasurfaces, which consist of sub-wavelength-spaced nanostructures with micron-scale thickness, have the potential to serve as ultralight and thin transmissive optics if they can be fabricated at scale over large areas. Although many early demonstrations of metasurfaces and binary subwavelength diffractive optics utilized electron-beam (e-beam) lithography[1–4] or other serial point-by-point fabrication techniques, such as two-photon optical lithography[5–7], which led to skepticism about their practical scalability, metasurfaces are now produced at larger scales using modern semiconductor foundry photolithography[8–10] and nanoimprinting techniques[11–13], alleviating such scaling concerns.

Increasing the metasurface diameter to accommodate large-aperture applications, however, does have physical and economic manufacturing limitations, as the number of required nanostructures scales as the square of the radius. E-beam lithography, which involves the serial writing of each nanostructure at high spatial resolution, is not a practical method for producing metasurface lenses (metalenses) larger than several centimeters in diameter. Even with state-of-the-art tools equipped with ultrafast 100 MHz-scale e-beam oscillators and multi-beam settings[14], the fabrication process takes several hours per sample, rendering it inefficient for large-scale production. Recent developments in e-beam writing techniques using variable-shaped beam (VSB) and cell-projection (CP) allow repeated metasurface structures to be written over a 280 mm diameter circle in 36 hours.[15] Nanoimprint lithography (NIL) is a viable technique for mass-producing metasurface optics or continuous metasurface films using metasurface patterns fabricated using e-beam lithography as molds[11,12,16], but the size of the mold itself is still limited by e-beam lithography limits, which directly restrict the maximum area of an individual NIL-made metasurface. A list of recent notable developments in scaling the diameter of metalenses is provided in Table S1.

The use of modern semiconductor foundry technology, which has supported the exponential growth in transistor density in integrated circuits since 1965[17], has been considered one of the most promising approaches for mass-producing metalenses. Previous works have demonstrated that *i*-line and deep-ultraviolet (DUV) projection lithography techniques, which enable high throughput nanofabrication by optically projecting the desired patterns onto a photoresist film, can be used to create centimeter-scale metalenses[8,9,11,18]. However, such lithography tools have an exposure size limit of around 20-30 mm, precluding single-shot fabrication of larger area metasurfaces (Fig. S1(a)).

Here, we experimentally demonstrate and characterize of a sub-meter scale (*i.e.*, 1/10[th] of a meter) large diameter metalens operating at visible wavelengths, fabricated using fully CMOS compatible process and materials. The metalens is manufactured using DUV projection lithography and has an area that exceeds the single shot exposure size limit of the lithography tool. We do this by stitching multiple exposure fields using different photolithography reticles in an exposure cycle. The metalens is polarization-insensitive, has a 100 mm diameter and 150 mm focal length for 632.8 nm wavelength light, corresponding to a numerical aperture (NA) of 0.32 or an *f*-number of 1.5. We choose fused silica ($SiO_2$) as the sole constituent material for this monolithic metalens not only for its low absorption in the visible and CMOS process compatibility but also for its high laser-induced damage threshold, which is suitable for high-energy applications. Also, the thermal robustness enables various coatings (*e.g.*, anti-reflection or anti-fouling coats) to be deposited directly onto the metalens or its backside, which is an essential requirement for such metalenses to be utilized in versatile astronomical applications. We show that this photolithography-based manufacturing approach, which was previously limited to the infrared spectrum[19–21] and to silicon wafers, can be extended to the visible spectrum and to more robust materials. The metalens' imaging quality is evaluated through its point-spread-function (PSF), modulation transfer function (MTF), focusing efficiency, and transmitted wavefront error via full-aperture interferometry. Then, we further simulate various fabrication error scenarios and quantify their individual impacts on wavefront and imaging quality. These simulations demonstrate that large area metasurfaces fabricated with multiple exposures and field stitching face fundamentally different aberration and efficiency challenges as compared to single-shot-photolithography metalenses and other ground refractive elements. We also show the suitability of the meta-optical device in extreme environments by evaluating its optical performance after thermal shock and temperature cycling across a 400°C range. Finally, we illustrate the remote and high-throughput imaging capabilities of the large diameter metalens by photographing celestial objects (the Sun, the Moon, and an emission nebula) in the visible[22,23], using the 100 mm diameter metalens as the only imaging lens.

## Results:

## Metalens design

Metasurfaces comprise nanostructures that impart a designed local wavefront or polarization transformation to incident light and can demonstrate unconventional functionality and performance with a very thin form factor that is difficult to achieve with conventional bulk optics. For instance, full Stokes polarimetry imaging has been realized with a single metasurface coupled to a commercial camera sensor[24–26], and an ultrathin perforated membrane metalens has been used to focus extreme ultraviolet radiation in transmission[27]. With precision design and manufacturing, metalenses have achieved diffraction-limited, high-efficiency performance[28,29], and attained broadband aberration correction when used in conjunction with existing refractive optics[30,31].

To design a 100 mm diameter metalens in a manner compatible with the fabrication workflow of existing semiconductor chip manufacturing tools, we first divide the 100 mm diameter region into 25 square sections (5×5 square array). Each section occupies a 20 × 20 mm square area (Fig. S1(b)), which is smaller than the exposure size limit of the DUV lithography tool used in this study (22 × 22 mm). We intentionally choose an odd number of arrays to ensure that the center of the metalens coincides with the center of a discretized section. This prevents unwanted scattering of light along the optic axis that may occur due to the stitching of the discretized fields near the metalens center, which may degrade imaging quality. Since the metalens is rotationally symmetric, the full 25 sections can be represented with just seven unique sections, six of which are repeated at four rotation angles (0°, 90°, 180°, and 270°) and one which is located at the metalens' center. Therefore, the 100 mm diameter metalens can be fabricated with 7 unique reticles (*i.e.*, DUV photomasks).

The metalens is composed of fused silica nanopillars with diameters ranging from 250 nm to 600 nm which have a constant edge-to-edge spacing of 250 nm and a height of 1.5 µm. This keeps both the pattern and the gap dimensions above the 200 nm single-exposure feature size limit of the DUV lithography tool used (ASML PAS 5500/300C DUV Wafer Stepper, λ=248 nm). The simulated transmitted phases and amplitudes versus the nanopillar diameter are plotted in Fig.

S2(a), which demonstrates phase coverage between $\pi/2$ and $3\pi/2$ radians. Although full 0 to $2\pi$ radian phase coverage can be achieved if the nanopillar height is increased to 2.1 μm, it is accompanied by a significant decrease in manufacturability and reduced structural integrity due to the increased structure width-to-height aspect ratio. We have thus limited the nanopillars to a maximum aspect ratio of 1:6. We have designated the empty area (*i.e.*, absence of a nanopillar) as a zero-phase element and included such empty areas in the design of the metalens to compensate for the limited library phase coverage. This approach maintains focal spot quality but reduces the focusing efficiency[8]. The diameter profile along the metalens radial direction is determined by selecting the nanopillar element at each radial position $r$ with a transmitted phase $\varphi(r)$ that most closely matches the desired ideal hyperbolic focusing profile[32,33]:

$$\varphi(r) = -\frac{2\pi}{\lambda_d}\left(\sqrt{r^2 + f^2} - f\right) + \varphi(0), \qquad (1)$$

where $\lambda_d$ is the design wavelength, $f$ is the focal length of the metalens, and $\varphi(0)$ is the phase at the center of the metalens. When Eq. 1 is satisfied, the transmitted light from each point on the metalens interferes constructively at the focal point; this is equivalent to having a spherical wavefront that converges at the focal point. We set $\varphi(0)$ to that of the transmitted phase of the nanopillar with the largest diameter in the library (600 nm) so that the phase wrapping zone transition between the largest and smallest nanostructures is located the furthest from the optical axis. The nanopillars at each radial position are placed uniformly across the azimuthal direction with a constant edge-to-edge spacing of 250 nm. The full metalens design comprises 18.7 billion nanopillars. The azimuthal spacing between nanopillars is slightly larger than 250 nm close to the metalens center due to the larger quantization error associated with the small integer number of nanopillars; however, such changes to the edge-to-edge gap have minimal effect on the transmitted phase from each nanostructure (see Fig. S2(b)). The maximum phase error resulting from such edge-to-edge spacing variation is 0.26 radians or 0.04 $\lambda$, which is below the Maréchal criterion of 0.25 $\lambda$ for achieving diffraction-limited performance[33–35].

The seven individual stitching sections of the metalens are indexed as reticles 1 to 7 (Fig. S1(b)), respectively. The center three reticles (reticles 1-3) were made by an industry-grade photomask manufacturer. The outer four reticles (reticles 4-7) were made in-house using a laser photomask writer with a larger critical feature size compared to that of the photomask manufacturer (Fig. S1(c)). Each reticle contains a 4-times magnified image of a 20 × 20 mm square section of the metalens, to match the DUV projection lithography tool demagnification ratio (Fig. S1(d)). The detailed process of the metalens design file generation, as well as the strategies employed to reduce file size and write times, can be found in the Supplementary Information.

As the first fabrication step, we coat a 150 mm (6-inch) fused silica wafer with a 150 nm-thick aluminum (Al) film (Fig. 1(a)), anti-reflective coating (ARC), and positive DUV photoresist layers (Fig. S3(a)). Then we create global alignment marks for the DUV lithography process. The alignment marks (*i.e.*, fiducials) are positioned and oriented such that the four alignment marks remain at the same location when the wafer is rotated by 0°, 90°, 180°, and 270° (Figs. S3(b), S3(c)). The patterned alignment marks in the photoresist layer are transferred into the aluminum film using wet etching, then both the ARC and photoresist layers are stripped with a downstream oxygen plasma ashing process.

We then coat the wafer with another ARC layer and a negative DUV photoresist layer, which serves as the etch mask for the metalens pattern creation in the Al layer. We opt to use a negative photoresist due to its advantageous process window compared to positive photoresist when creating isolated structures: overexposure of positive resist can cause shrinkage or delamination of the resist patterns, while patterns on negative resist simply become enlarged. As shown in Fig. 1(b), the wafer is first loaded into the DUV lithography tool at 0° orientation, and reticles 1-7 are aligned and exposed. The alignment error of the used stepper lithography tool is less than 45 nm, which is about 7 % of the target wavelength and thus not expected to significantly distort the transmitted wavefront. State-of-the-art photolithography tools can achieve overlay errors down to 1 nm. The wafer is then rotated to 90°, and reticles 2-7 are exposed. The same process is repeated for the 180° and 270° wafer orientations, followed by baking and development (Fig. S4(a)). The alignment and exposure of the entire 100 mm metalens is fully automated and takes less than 20 minutes per wafer.

After the metalens pattern is formed in the negative photoresist, it is first transferred to the ARC layer using an Ar/$O_2$ reactive ion etch (RIE) (Fig. S4(b)) and then transferred into the Al layer using Ar/$Cl_2$ inductively coupled plasma reactive ion etching (ICP-RIE), as shown in Fig. 1(c). $SiO_2$ is thermodynamically protected against chemical etching by $Cl_2$ plasma[8] and acts as an etch stop layer. Both the ARC and resist layers are then stripped using a downstream oxygen plasma ashing process (Fig. S4c), which does not significantly affect the patterned Al film. Using the patterned Al as a hard etch mask, we then vertically etch into the fused silica substrate with ion-enhanced inhibitor etching using an optimized octafluoropropane ($C_3F_8$) ICP-RIE process (Fig. S5(a)) until the etch depth reaches 1.5 μm. The measured uniformity of the etching speed across the 100 mm diameter region is provided in Fig. S6, which shows that the maximum resulting etch depth difference across the entire 100 mm diameter metalens is approximately 80 nm, corresponding to 5 % of the target pillar height. This height deviation is associated with a phase error smaller than the quarter-wave deviation acceptable under Maréchal criterion (Fig. S2(d)). Once the desired etch depth is reached, the residual Al film on the top of the nanopillars is selectively etched away using an Ar/$Cl_2$ ICP-RIE process, leaving the fused silica nanopillars. More details on the fabrication process can be found in the Materials and methods section.

## Optical performance characterization

A photograph and a tilted scanning electron microscope (SEM) image of the 100 mm diameter metalens are shown in Fig. 2(a) and 2(b), respectively. The vertical and smooth sidewalls of the etched fused silica nanopillars are visible, which is a notable improvement from previous fused silica metalens work that resulted in tapered and rough sidewalls[8]. More detailed SEM images of the metalens are provided in Supplementary Fig. S7. Figs. 2(c) and 2(d) present a side-by-side comparison between the fabricated metalens and an off-the-shelf plano-convex refractive lens made of N-BK7 glass with MgF$_2$ anti-reflection coating (#19-904, *Edmund Optics*) with a similar diameter ($D$=100 mm) and focal length ($f$=150 mm at λ=587.6 nm). The metalens including the substrate is 42 times thinner (0.5 mm vs. 21 mm) and 16.5 times lighter (14.6 g vs. 242.2 g) than the refractive lens counterpart.

To evaluate the focusing quality of the metalens, we illuminate the pristine backside of the metalens with the expanded and collimated beam of a Helium-Neon laser (λ=632.8 *nm*). We then acquire the focal profile along the optical axis separately using a horizontal microscope (Fig. S8) and a point-source microscope attached to a coordinate measuring machine (Fig. S9). The measured distance to the focal plane from the metalens is 149.97±0.18 mm, which agrees well with the designed focal length of 150 mm (Fig. S9). Figs. 3(a) and 3(b) show a transverse and a longitudinal cut of the measured Point-Spread Function (PSF). For comparison, the simulated transverse and longitudinal focusing profiles for a diffraction-limited metalens with NA=0.32 are provided in Figs. 3(c) and 3(d).

The Strehl ratio of the fabricated metalens is 0.6. Furthermore, the diameter of the first ring of minimum intensity around the focal point of the metalens is 3.4 μm, whereas the ideal Airy disk diameter for a diffraction-limited lens with NA=0.32 would be 2.3 μm, indicating the presence of aberrations in the metalens. The measured PSF, degraded by the wavefront aberration, resembles that of a lens with an effective NA of 0.2, which corresponds to a lens with a diameter of 60 mm and a focal length of 150 mm that has an Airy disk diameter of 3.86 μm (Fig. 3(e)). This similarity becomes more apparent when the 2D modulation transfer functions (MTFs) are calculated from the PSF images and compared with the ideal MTFs for lens NAs of 0.32 and 0.2, as shown in Fig. 3(f).

The primary cause of the aberration is the reticle quality difference between the industry-grade center sections (reticles 1-3) and the lab-made outer sections (reticles 4-7). The SEM images in Fig. S7 show a small nanopillar diameter mismatch between the two different reticle sources. This indicates that the nanopillar diameters in the outer sections were not correctly fabricated. Such a diameter mismatch results in a phase discontinuity between the inner and outer exposure fields. The more significant fabrication error, however, is that in the inner region, all nanopillars are present as per the design, but in the outer regions, the small diameter pillars are missing due to those features not being resolved in the reticles fabricated with the lower resolution in-house laser writer. Both errors are avoidable by using reticles from a single manufacturing source.

The effect of the missing pillars on the PSF becomes more apparent when the transmitted phase and the diffracted amplitude information are decoupled using interferometry and diffraction intensity imaging. The deviation of the transmitted wavefront phase of the metalens from the ideal, *i.e.*, wavefront aberration function (WAF), is obtained using full-aperture interferometry. For this, a commercial Fizeau interferometer (Zygo VeriFire[TM] ATZ, *Zygo Corporation*) is used in a double-pass nulling interferometry configuration, in which the metalens-focused beam is reflected back by a return sphere (*i.e.*, high-accuracy spherical ball), and produces an interferogram at the interferometer aperture (Fig. S10). Due to the double-pass configuration, the metalens WAF is half of the measured WAF. More details of the interferometric procedure are provided in the Supplementary Information. The map of the metalens WAF is shown in Fig. 3(g), with the root-mean-square (RMS) error over the entire 100 mm aperture of 0.08 λ and a peak-to-valley (PV) error of 0.64 λ (after removing interferometry alignment terms, *i.e.*, piston, tip/tilt, and power). The inner 60 × 60 mm section, fabricated with industry-grade photomasks, displays lower RMS and PV values of 0.05 λ and 0.28 λ, respectively. These values indicate that the industry-grade fabrication technique can attain phase error values close to the diffraction-limit criterion of an RMS below 0.075 λ and a PV below 0.25 λ[33,34,36]. The fitted Zernike polynomial coefficient values to the WAF are provided in Fig. S11 and Table S2, which indicate that the optical aberrations are dominated by 2nd- and 4th-degree contributions. The expected tilt dependence of the as-manufactured metalens' wavefront (*i.e.*, coma) was also successfully evaluated and confirmed interferometrically (Fig. S15). The RMS wavefront error rises to 1 λ for tilts around 10 arcseconds.

The diffracted intensity after the metalens (Fig. 3(h)), *i.e.*, light contributing to the focus, shows a stark difference between the efficiency of the inner (reticles 1-3) and outer (reticles 4-7) lens sections. The diffracted image is captured at a plane displaced 11 mm axially away from the focal plane, in the direction towards the metalens. We find that 86% of the power that contributes to the focus comes from the inner region, which comprises only 46% of the total metalens area, causing the focal spot to resemble that of a square 60 × 60 mm lens instead. As discussed earlier, the reduced diffraction intensity in the outer region is mainly due to missing small diameter pillars in the in-house fabricated photomasks. The measured focusing efficiency of the entire metalens, relative to that of the equivalent off-the-shelf $MgF_2$-coated plano-convex refractive lens in Fig. 2(d), is 40.4%. However, when only the inner region is profiled, the focusing efficiency rises to 63.1%, which agrees with the diffracted intensity profile. Thus, the reticle quality difference across each exposure field is the primary cause of the aberrated PSF.

## Fabrication error tolerance study

Exposure field-linked fabrication errors are a unique problem for metasurface optical elements made with CMOS-compatible manufacturing processes, since such rectilinear imperfections do not exist for traditional optics manufacturing methods such as diamond turning or other mechanical grinding and polishing techniques. To fully characterize such fabrication errors on the imaging quality degradation and their corresponding tolerances, we conduct a simulation-based study quantifying the following effects: (1) unresolved nanopillars below certain diameters, (2) a uniform shift in nanopillar diameters, and (3) a uniform shift in nanopillar height. Effects (1) and (2) can occur during photomask manufacturing or through shifts in exposure and development conditions, and (3) can occur if the nanopillars are either under- or over-etched. We consider the effects of these errors when they are applied to each reticle-linked section individually and across the inner and outer sections (Fig. S12).

Since full-lens simulation of 18.7 billion nanostructures is impractical on full-wave Maxwell equation solvers, we employ the locally periodic assumption[37] and take the transmitted field to be equal to the stitched fields from individual nanopillars. We numerically discretize the metalens into 100,000 annular rings, each divided into 100 sections evenly in the azimuthal direction, resulting in a total of 10 million arc sections. This discretization choice allows for fine resolution of the rapidly-varying radial transmission behavior without expending additional elements to resolve the slowly-varying azimuthal behavior. We then propagate the transmitted electromagnetic field from each arc section toward the focal plane using a vectorial propagator[38], weighting each arc section with its area. The focal plane complex fields are used to calculate the MTF, Strehl ratio (volume under the 2D MTF relative to that of the diffraction limit), and focusing efficiency (fraction of incident power that passes through 3 Airy disk diameters centered on the optic axis). Further details of the simulation are provided in the Supplementary Information.

Fig. 4(a)-(c) present the simulated wavefront of a metalens where nanopillars smaller than a given threshold diameter are unresolved for (a) all 7 sections, (b) only the inner sections (reticles 1-3), and (c) only the outer sections (reticles 4-7), respectively. The simulated Strehl ratios and focusing efficiencies of each case with respect to the minimum fabricated pillar diameters are shown in Fig. 4(d-e), respectively. The Strehl ratio degrades when either the inner or outer sections' smaller pillars are missing but remains near unity for the case when the pillars below a given diameter are equally missing over all sections. This reduction in the Strehl ratio occurs due to incomplete constructive interference near the focus due to a mismatch in diffracted intensity contributions from different parts of the lens, which broadens the transverse peak. However, when all sections have similar amounts of missing pillars, such diffracted intensity variations are eliminated, and one obtains diffraction-limited focusing. In comparison, the focusing efficiency in all three cases decreases monotonically as the minimum fabricated pillar diameter increases, since more light passes straight through the metalens and is not deflected to the focus. This is consistent with previous findings, in which metalenses with unfabricated smaller structures were still able to produce diffraction-limited focusing albeit at reduced efficiencies[8]. The MTFs for the three scenarios when the minimum fabricated nanopillar diameter is 550 nm are shown in Fig. 4(f). The MTF curves are normalized to unity at their zero frequency values, respectively. When all sections have missing smaller diameter pillars, the MTF remains diffraction-limited. When the inner

sections have missing pillars, the MTF at lower spatial frequencies is reduced, and vice versa for outer sections having unfabricated pillars. More detailed plots of the MTF behavior under other missing pillar conditions are provided in Fig. S13(a-c).

For fabrication errors other than section-linked missing nanopillars, the Strehl ratio remains diffraction-limited although the focusing efficiency decreases. The calculated Strehl ratios and focusing efficiencies resulting from nanopillar diameter shifts are provided in Fig. S13(d-e), respectively. The metalens retains diffraction-limited performance for nanopillar diameter shifts less than 100 nm. Similarly, an error in nanopillar height, which can occur during the final $SiO_2$ etching process, does not significantly affect the Strehl ratio (Fig. S13(f)) but reduces the focusing efficiency. This retention of high Strehl ratio focusing arises because the phase relationship between nanopillars of different diameters remains largely unchanged with respect to systematic diameter and height variations (Fig. S2(a,d)). However, as the relative phase between the empty areas and the nanopillars shifts, the efficiency is reduced due to out-of-phase interference at the focal plane.

Beyond fabrication errors, we investigate the role of imperfections in the measurement apparatus, which arise easily due to the challenges of aligning and characterizing large aperture optical elements. If the illuminating light is non-uniform, for example, with a Gaussian intensity profile (Fig. S14(a)), the measured PSF becomes aberrated and the Strehl ratio falls below unity, even for a perfectly fabricated metalens (Fig. S14(b)). However, the Gaussian width of the illumination must be substantially smaller than the lens diameter (20 mm Gaussian width) to reduce the Strehl ratio below the diffraction-limited level of 0.8. Tilts in the illuminating field (*i.e.*, off-axis illumination) can also introduce field-dependent aberrations in the PSF. Fig. S16 exhibits the simulated reduction in measured Strehl ratio as a function of incident angle tilts. The illuminating field must be normally incident to within 10 arc seconds to provide an accurate measurement of the metalens focusing quality, which agrees with interferometric measurements shown in Fig. S15. Our PSF characterization setup (Fig. S8) has an illumination uniformity and tilt tolerance below these thresholds.

Table S3 summarizes the main conclusions of the fabrication and measurement error simulation study. Only a mismatch between the unfabricated pillars in the inner and outer sections produces the Strehl ratio reduction observed in the experimental focal spot.

## Meta-imaging the cosmos

To illustrate the imaging performance of the fabricated metalens, we build a wide field-of-view (*i.e.*, ~0.6°) meta-astrophotography apparatus using only the 100 mm diameter metalens (without any other optical lenses or mirrors), a narrowband color filter, and a cooled CMOS imaging sensor placed at the metalens focal plane (Fig. 5(a)). The distance between the metalens and the imaging sensor is controlled by a helical focuser and the apparatus is enclosed to eliminate stray light. We note that this is not a telescope, as there is no second lens and no magnification involved. The meta-astrophotography system is mounted on an equatorial mount with a guide-scope for real-time sky tracking, and is used to capture images of sunspots on the Sun (with a 633 nm bandpass filter, Fig. 5(b)), the North America Nebula (with a 656.28 nm Hydrogen alpha bandpass filter, Fig. 5(c)), and the Moon (with a 633 nm bandpass filter, Fig. 5(d)). All images were captured in Cambridge, Massachusetts, USA. Higher resolution images and details of the image acquisition methods are provided in Figs. S17, S18, and S19, respectively. More photographs taken with the 100 mm diameter metalens are provided in Fig. S26.

Such imaging of celestial objects in the visible illustrates that the 100 mm diameter all-glass metalens is suitable for remote imaging applications. To deploy meta-optics in remote imaging platforms such as high-altitude UAV platforms, low-earth orbit satellites, or off-world exploration spacecraft and vehicles, the nanostructures must withstand extreme environmental stresses such as large temperature swings, cosmic radiation, and vibration[39–41]. To test whether the all-glass metalens can survive harsh environments, we devised a thermal shock and heat stress cycling test resembling that of the United States Department of Defense Test Method Standard (MIL-STD-883F). The subject in testing was cyclically moved between a cold reservoir (liquid nitrogen, -195.8 °C) and a hot reservoir (hot plate, 200 °C). The subject remained in each thermal reservoir for 10 minutes to reach thermal equilibrium, while the sample transfer time between the two thermal baths was less than 5 seconds to induce thermal shock. More details of the devised test method are provided in Fig. S20. After performing the stress test on a 10 mm diameter all-glass metalens[8] for 10 cycles and returning the sample to room temperature, we do not observe significant change in the optical performance (Fig. S21) nor physical damage after 15 cycles (Fig. S22). This is due to fused silica having a low-level of impurities and also a near-zero thermal expansion coefficient ($\alpha \approx 0.5 \times 10^{-6}/K$)[42] which would result in an approximately 0.01 % shift of the metalens radius over the 400 °C temperature range. As the thermo-optic coefficient ($dn/dT$) of fused silica is also very low ($< 10^{-6}/K$)[43], we expect that the change in the transmitted wavefront with respect to the temperature will be low as well. Therefore, we anticipate the impact of thermal variations on the optical performance of all-glass metalens to be insignificant within the given test temperature range. The metalens also does not exhibit noticeable physical damage under the vibrational stress induced by immersing the sample in an ultrasonication bath for 20 minutes (Fig. S23). These results suggest that the all-glass metasurfaces can survive extreme environment conditions and therefore is well-positioned for space applications requiring launch survival. Similar tests conducted on $TiO_2$ metasurfaces comprising 700 nm tall nanopillars[44] also show promising results as well (Fig. S24, S25). The thermal resilience also allows various optical coatings to be applied onto the metalens after fabrication, since such modifications require special thermal environments during coating.

**Discussion:**

In summary, we demonstrate the fabrication of an all-glass 100 mm diameter metalens capable of operating in the visible wavelength range using DUV lithography, surpassing the tool's exposure size limit. The diameter of the metalens can be further increased to about 290 mm, as 300 mm diameter wafers and corresponding CMOS foundry tools become increasingly available in the industry. We also show that in addition to PSF measurements, full-aperture interferometry offers valuable insights into the optical wavefront-based characterization of metalenses. This technique effectively decouples the phase and amplitude error information, which identifies a unique challenge for flat optic elements fabricated by stitching multiple exposure fields. Through simulation-based study, we also show the effects of various possible exposure field-linked fabrication errors on the imaging quality of the metalens, which provide tolerance windows for the manufacturing processes. Additionally, we demonstrate that the single metalens is capable of imaging celestial objects in the visible wavelength range.

The efficiency of the presented metalens can be improved by using all-industrial grade reticles and higher resolution manufacturing techniques such as immersion DUV lithography ($\lambda$=193 nm), so that even smaller diameter nanopillars can be fabricated[9,12]. Dispersion engineering with anisotropic pillar shapes can also be employed to achieve broadband high efficiency[28]. As fused silica has relatively high laser-induced damage threshold (LIDT) compared to most optical glasses[45], we anticipate that the all-glass metasurface platform will be useful in versatile coating options and high-power laser applications not only for focusing but also for polarization manipulation[26] and pulse compression[46,47]. Furthermore, their resilience under extreme environmental conditions highlights their suitability for remote imaging in harsh environments. Additionally, the demonstrated fabrication process holds promise for creating large-diameter aberration-correcting meta-optics[30,31]. These meta-optics may replace or enhance optical components in existing multi-element optics systems, providing a path towards low-weight, high-performance, large-aperture imaging devices.

## Material and methods:

### <u>Fabrication</u> <u>process</u>

150 mm diameter, double side polished, 500 $\mu m$ thick JGS2 fused silica wafers (*WaferPro LLC*) are used in this study. During the fabrication process, the creation of a photoresist edge-bead, either by recoil or surface tension, can introduce unreliable fabrication conditions. This includes challenges such as difficulties in wafer height detection or achieving plasma etch uniformity at the edge of the wafer. In addition, some plasma etching tools mechanically clamp down on edge of the wafer during processing that the edge of the area should be avoided. Therefore, it is generally a good rule of thumb to exclude a few millimeters around the rim of a wafer during wafer-based manufacturing. As conventional semiconductor foundries typically use 100 mm (4-in.), 150 mm (6-in.), 200 mm (8-in.), or 300 mm (12-in.) wafers, here we choose 150 mm diameter wafer as a substrate for a 100 mm diameter metalens.

The wafers are cleaned using a MOS clean process involving an organic cleaning step by immersing the wafers in 1:1:6 volumetric ratio $H_2O_2:NH_4OH:H_2O$ solution at 85 °C for 10 minutes, followed by an ionic clean step by immersing in 1:1:6 $H_2O_2:HCl:H_2O$ solution at 85 °C for 10 minutes. After each cleaning step, the wafers are thoroughly rinsed in deionized (DI) water bath by filling and dumping the rinse bath 3 times. The MOS cleaned wafers are then spin-rinsed and dried (Superclean 1600 Spin Rinse Dryer, *Verteq*). The 150 nm thick Al film is deposited on the cleaned wafers using an e-beam evaporator with planetary wafer holders (CHA MARK 50 E-beam Evaporator, *CHA Industries*).

The Al-coated wafers are first spin-coated with 62 nm thick DUV anti-reflection coating (ARC) layer (DUV 42P, *Brewer Science*), followed by spin-coating of 600 nm thick positive DUV resist (UV210-0.6, *Shipley*). The alignment marks are exposed at the designated positions as shown in Fig. S3 using a DUV projection lithography tool (PAS 5500/300C DUV Stepper, *ASML*). The photoresist film is then post-exposure baked (PEB) and developed using AZ 726 MIF developer. We note that all spin-coating, baking, and developing process are performed with a SUSS MicroTech Gamma automatic coat-develop tool for consistent fabrication results. The developed alignment marks are then transferred to the ARC layer using $Ar/O_2$ reactive ion etching (RIE, PlasmaLab80Plus, *Oxford Instruments*), exposing the underlying Al film. Then, the alignment markers are wet-etched into the Al film to a depth of approximately 120 nm to meet the requirement for phase-contrast detection system in the DUV stepper system. The resist and ARC layer are then stripped using downstream oxygen plasma ashing (YES EcoClean, *Yield Engineering Systems*).

The wafer with alignment marks (*i.e.*, fiducials) etched into Al layer is then coated with 62 nm thick DUV ARC layer, followed by 500 nm thick negative DUV resist (UVN2300, *Dow Electronic Materials*). The 100 mm diameter metalens pattern is exposed onto the wafer with the DUV stepper system as depicted in Fig. S1, where reticles 1-3 are fabricated by an industry-grade photomask manufacturer and the reticles 4-7 are prepared on chrome coated reticle plates using the DWL2000 Laser Writer (*Heidelberg Instruments*) followed by developing and chrome wet-etching with the HMP900 Mask Processing System *(Hamatech)*. We note that the DUV photoresists used in this study are chemically amplified resists, and hence it is important to minimize the time between the exposures of the first section and the last to achieve the desired pattern sizes throughout the wafer. In detail, the photoresists used in this study consist of compounds that generate acid upon photoactivation. The shape and size of the developed photoresist profile is determined by the acid-induced reaction that occurs during the post-exposure bake step. Since the photo-generated acid diffuses into the photoresist film over time, any significant delay between exposures of the discretized sections can lead to differences in the feature dimensions across those sections. Therefore, reducing the number of sections to be aligned and exposed, or the overall time of exposure, can help ensure consistency of the feature sizes between the sections. The exposed resist layer then goes though PEB and development processes using AZ 726 MIF developer solution and the SUSS MicroTech Gamma automatic coat-develop tool. The metalens pattern is then transferred to the ARC layer using $Ar/O_2$ RIE (PlasmaLab80Plus, *Oxford Instruments*).

Using the patterned resist and ARC layer as an etch mask, the metalens pattern is then transferred to Al film using $Ar/Cl_2$ inductively coupled plasma RIE (ICP-RIE, PlasmaPro 100 Cobra 300, *Oxford Instruments*). The resist and ARC layer are then stripped using downstream plasma ashing (Matrix 105, *Matrix Integrated Systems*), leaving only the patterned Al on the wafer. Then, the 1.5 μm tall, fused silica nanopillars are formed by vertically etching into the fused silica substrate using $C_3F_8$ ICP-RIE (NLD-570, ULVAC), using the patterned Al as an etch mask. In detail, when $C_3F_8$ is introduced in the etch chamber and is dissociated into plasma, a thin film of fluorocarbon deposits on the exposed area of the substrate, creating an etch inhibitor layer that prohibits fluoride ions from chemically etching $SiO_2$ (Fig. S5(a)). The fluoride ions, which are accelerated toward the surface of the substrate, physically bombard and break the fluorocarbon film, and chemically etch the $SiO_2$ underneath anisotropically. By balancing the growth rate of the fluorocarbon film and the etch speed with temperature, one can control the sidewall tapering angles during the $SiO_2$ etch (Fig. S5(b-i)). Details of the vertical etch process is provided in Fig. S5 and S6. Then, the residual Al on top of each nanopillar is removed using selective etch with $Ar/Cl_2$ ICP-RIE (PlasmaPro 100 Cobra 300, *Oxford Instruments*).

## **Acknowledgements**

The authors would like to thank Garry Bordonaro, John Treichler, Aaron Windsor, Jeremy Clark, and Chris Alpha for their generous help in using CNF facilities. The authors would also like to thank Dr. Rohith Chandrasekar and Lidan Zhang for insightful discussions and Dr. Seong Soon Jo for help with metalens wafer dicing. This work was supported by the Defense Advanced Research Projects Agency (DARPA) Grant No. HR00111810001. This work was performed in part at the Cornell NanoScale Science & Technology Facility (CNF), a member of the National Nanotechnology Coordinated Infrastructure (NNCI), which is supported by the National Science Foundation (Grant NNCI-1542081), and in part at the Harvard University Center for Nanoscale

Systems (CNS); a member of the National Nanotechnology Coordinated Infrastructure Network (NNCI), which is supported by the National Science Foundation under NSF award no. ECCS-2025158. The computations in this paper were run on the FASRC Cannon cluster supported by the FAS Division of Science Research Computing Group at Harvard University. S.W.D.L. is supported by A*STAR Singapore through the National Science Scholarship scheme.

## Author contributions

J.S.P. and F.C. conceived the study. J.S.P., S.W.D.L., and M.O. developed codes. S.W.D.L. and J.S.P. performed simulations and their analysis. J.S.P. fabricated the samples. J.S.P. and S.W.D.L. analyzed the fabrication results. J.S.P., S.W.D.L., H.K., K.K., D.K., J.L., A.M.U., and Z.L. performed metalens characterization and analyzed the data. J.S.P. and A.A. built the meta-astrophotography apparatus. A.A. performed the image acquisition of the celestial objects. S.W.D.L. and A.A. analyzed the celestial images. All authors contributed to the writing of the manuscript, discussed the results, and commented on the manuscript.

## Conflict of interest

The authors declare no competing interests.

## Data availability

Data of this study is available from the corresponding author upon reasonable request.

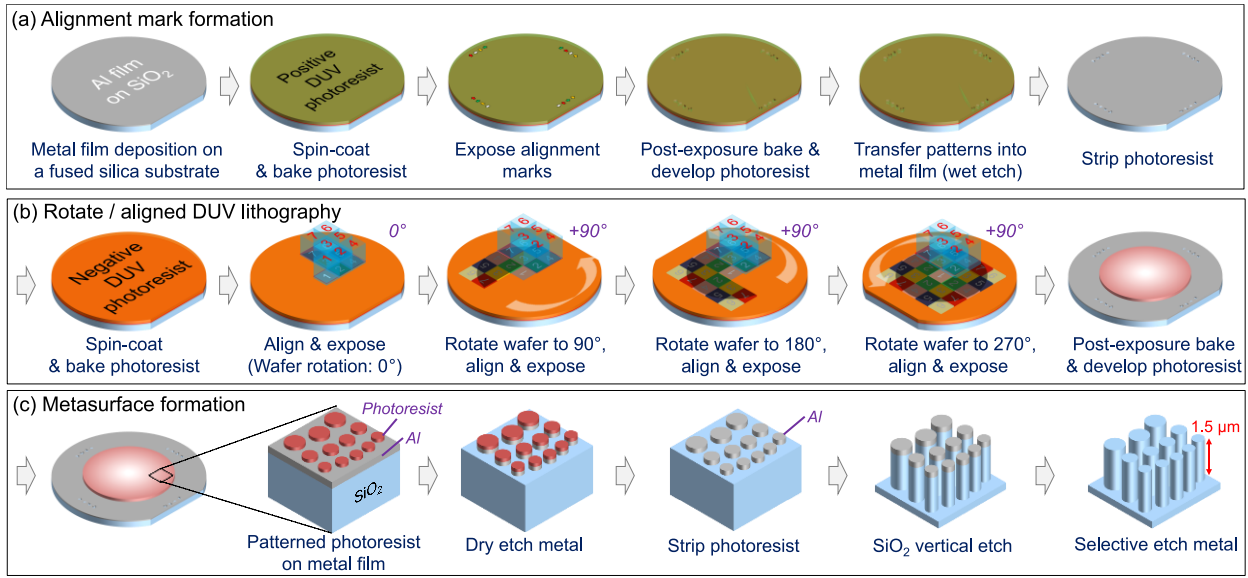

**Fig. 1. Fabrication process of the 100 mm diameter, all-glass metalens.** (a) Global alignment marks are formed by etching the alignment patterns into the deposited aluminum film on a fused silica substrate. (b) A 100 mm diameter metalens pattern, segmented into 25 sections, is formed on the aluminum film using a negative DUV photoresist with 7 photomasks. After exposing sections 1-7 at $0°$ wafer orientation with global alignment onto the negative DUV photoresist film, sections 2-7 are also aligned and exposed on the resist layer at $90°$, $180°$, and $270°$ wafer rotation, respectively. The exposed DUV photoresist is baked and developed, leaving the metalens pattern in the aluminum film. (c) Using the photoresist pattern as an etch mask, the metalens pattern is transferred to the aluminum film with an $Ar/Cl_2$ plasma etch. The photoresist layer is stripped using downstream oxygen plasma ashing. Using the patterned aluminum as etch mask, we perform a vertical $SiO_2$ etch with $C_3F_8$ plasma into the fused silica substrate until the desired pillar height of $1.5\ \mu m$ is reached. The residual Al film is removed using a $Cl_2$ plasma etch, leaving only the fused silica nanopillars.

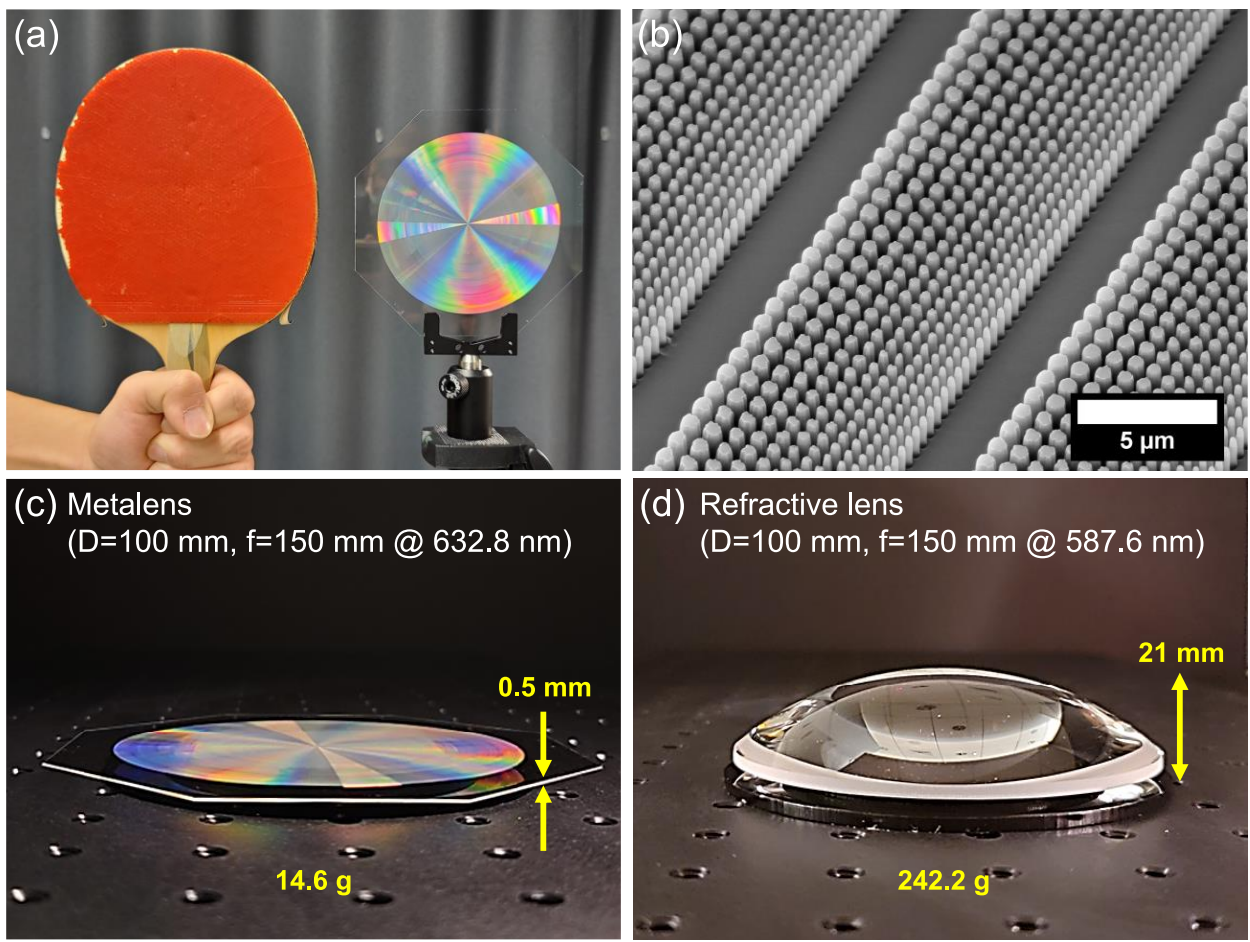

**Fig. 2. Photographs and SEM images of the 100 mm diameter, all-glass metalens.** (a) Photograph of the metalens compared with a table tennis racket. (b) SEM image of the fused silica nanopillars comprising the metalens. More SEM images are provided in Supplementary Information (Fig. S7). Photographs taken from the side of (c) the metalens ($f = 150\ mm$ at $\lambda = 632.8\ nm$) and (d) a plano-convex lens made of N-BK7 glass with a similar diameter (100 mm) and focal length ($f = 150\ mm$ at $\lambda = 587.6\ nm$, *Edmund Optics* #19-904). The thickness and weight of the metalens are 0.5 mm and 14.6 g, and those of the refractive lens are 21 mm and 242.2 g, respectively.

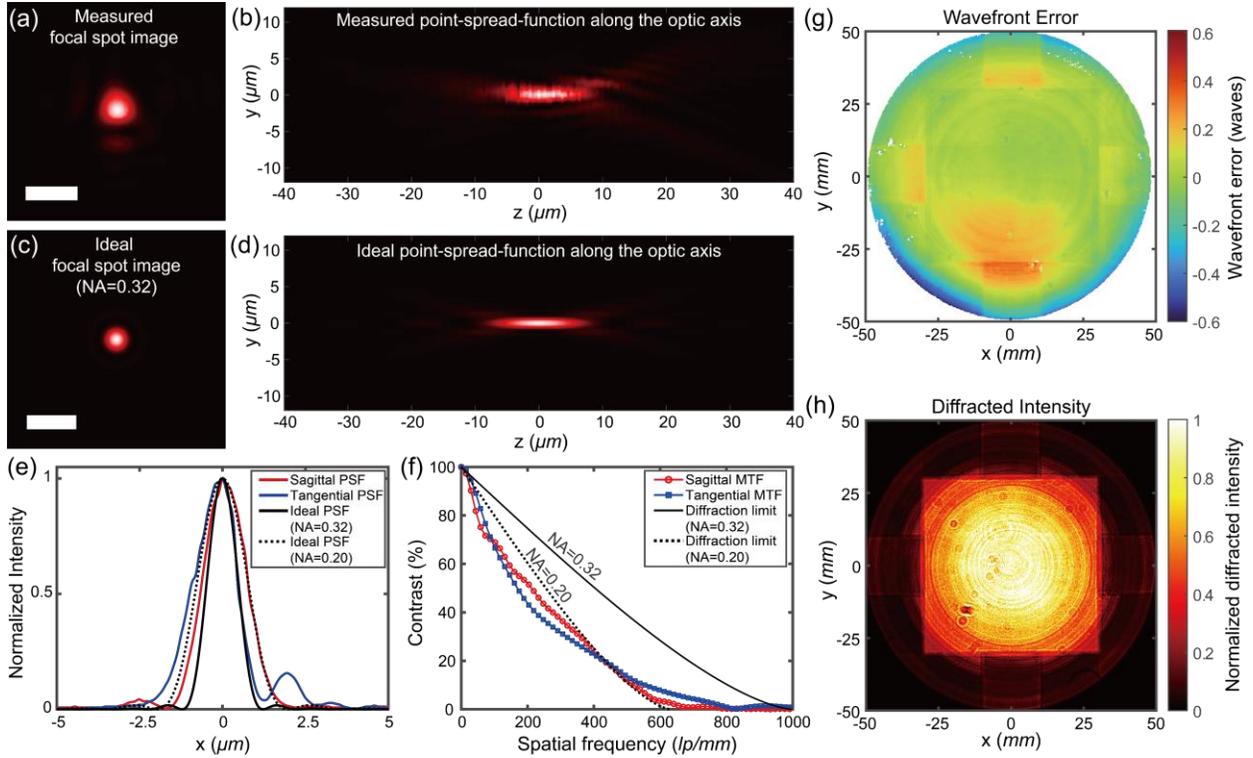

**Fig. 3. Optical characterization of the 100 mm diameter, all-glass metalens.** (a) Focal spot image of the fabricated metalens at the design wavelength ($\lambda = 632.8$ nm). Scale bar: 3 $\mu m$. (b) Point-Spread-Function (PSF) of the metalens along the optic axis. $z = 0$ $\mu m$ represents the focal plane ($f$=150 mm). (c) Simulated focal spot image of an ideal lens with NA=0.32 (scale bar: 3 $\mu m$), and (d) simulated PSF of an ideal lens along the optic axis. (e) Focusing profiles along the sagittal (vertical) and the tangential (horizontal) plane of the metalens' focus shown in (a) and simulated focusing profiles of diffraction-limited lenses with NAs of 0.32 (100 mm aperture, 150 mm focal length) and 0.20 (60 mm aperture, 150 mm focal length), respectively. (f) Modulation transfer function (MTF) of the metalens along the sagittal and the tangential direction, and diffraction-limited MTFs of lenses with NA of 0.32 and 0.20, respectively. (g) Measured wavefront error map of 100 mm diameter metalens, obtained from full-aperture interferometry. The interferometry alignment terms (*i.e.*, piston, tip, tilt, and power) are removed. (h) Intensity image of the 1st-order diffracted beam from the metalens, which corresponds to the intensity contribution from each point of metalens to the focal point. The diffracted intensity is higher in the central 6 × 6 cm region, at which industrial-grade photomasks were used (reticles 1-3), compared to the outer region at which lab-written photomasks were used (reticles 4-7). The mismatch in diffraction efficiency between the inner and outer regions causes a broadening of the observed focusing behavior from that of an ideal 100 mm diameter lens.

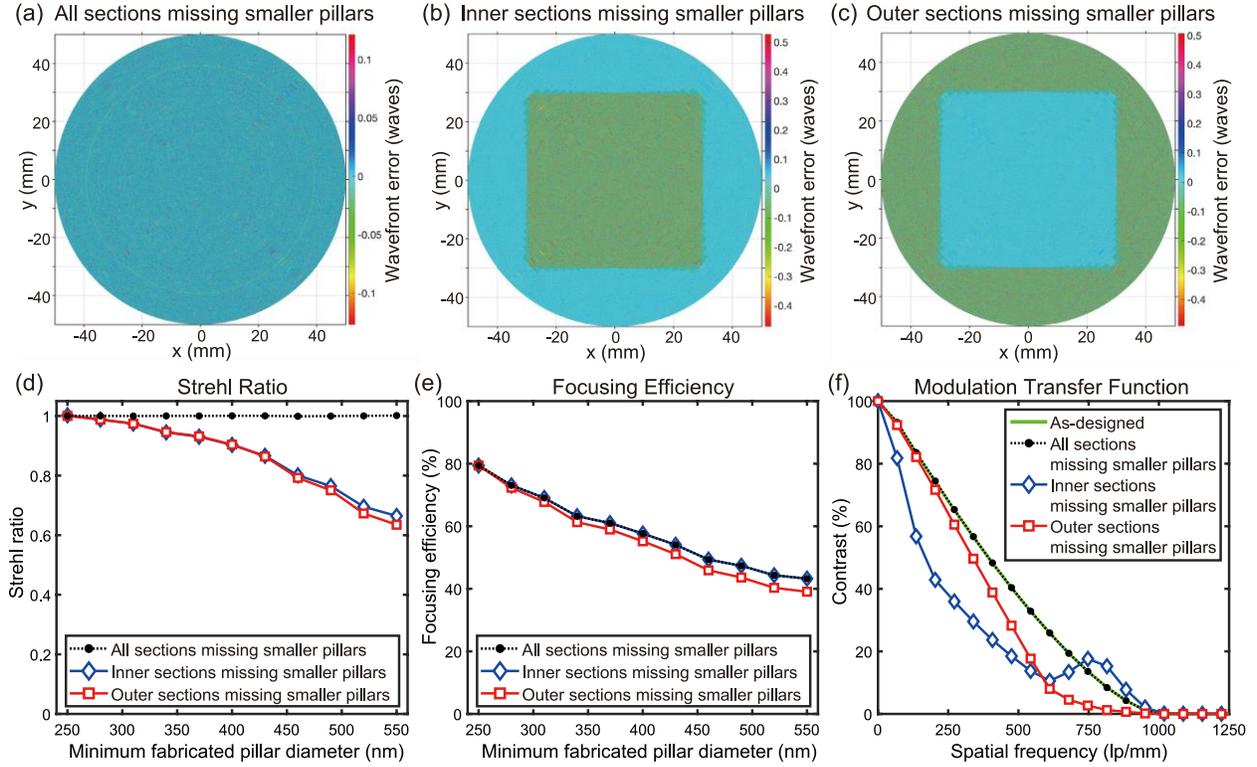

**Fig. 4. Simulation-based study on the effects of missing smaller pillars in different regions of metalens on its imaging quality and efficiency.** Simulated wavefront error of a metalens missing small pillars (a) in all sections (reticles 1-7), (b) in the inner sections only (reticles 1-3), and (c) in the outer sections only (reticles 4-7), respectively. The surface plots show the resulting wavefront error when the smallest fabricated nanopillar diameter is 550 nm. (d) Simulated Strehl ratio versus the minimum fabricated nanopillar diameter for all sections, inner sections only, and outer sections only, respectively. The Strehl ratio remains diffraction-limited when the nanopillars are missing uniformly across the entire metalens, while a mismatch in the loss of nanopillars result in poorer imaging quality due to the resulting wavefront error. (e) Simulated focusing efficiencies when all sections, inner sections only, and outer sections only are missing smaller pillars. All cases exhibit degradation of overall focusing efficiencies due missing smaller diffracting elements. (f) Normalized modulation transfer function (MTF) for each scenario of missing pillars when the smallest fabricated nanopillar diameter is 550 nm. The MTF contrast curves are normalized to their respective zero-frequency values.

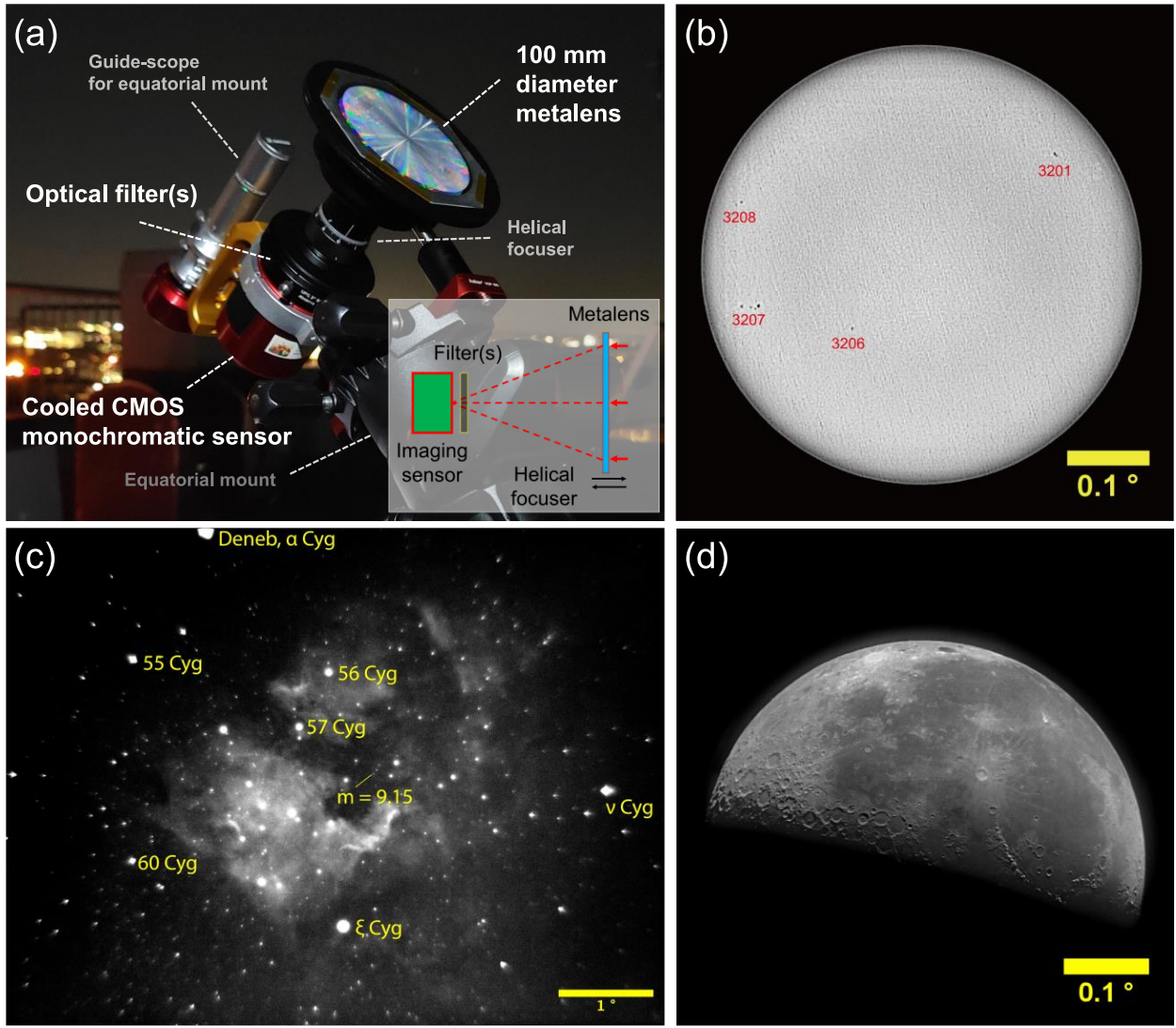

**Fig. 5. Imaging the heavens in the visible with 100 mm diameter wide field-of-view (~0.6°) meta-astrophotography apparatus.** (a) Photograph of the astro-imager comprising only a 100 mm diameter metalens, an exchangeable optical filter, and a cooled CMOS monochromatic sensor, mounted on an equatorial mount guided by a guide-scope. (b) Acquired image of the Sun with a neutral density filter (OD 3.0) and a 1 nm bandwidth bandpass filter centered at 632.8 nm. Image taken on Feb. 01, 2023, Cambridge, Massachusetts, USA. Identified sunspot group numbers are labeled. (c) Acquired image of the North America Nebula (NGC 7000, in the constellation Cygnus) with a 7 nm bandwidth $H$-$\alpha$ filter ($\lambda = 656.28$ nm). Image taken on May 13, 2022, Cambridge, Massachusetts, USA. Notable celestial objects are labeled. The imaging system can detect stars down to an apparent magnitude of 9.15. (d) Acquired image of the Moon at its last quarter phase, taken with a 1 nm bandwidth bandpass filter centered at 632.8 nm. Image taken on Aug. 18, 2022, Cambridge, Massachusetts, USA. High-resolution images are provided in the Supplementary Information.

Supplementary Information for

# All-glass 100 mm Diameter Visible Metalens for Imaging the Cosmos

**Authors:** Joon-Suh Park[1]*†, Soon Wei Daniel Lim[1]†, Arman Amirzhan[1], Hyukmo Kang[2], Karlene Karrfalt[2,3], Daewook Kim[2], Joel Leger[3], Augustine M. Urbas[3], Marcus Ossiander[1,4], Zhaoyi Li[1], and Federico Capasso[1]*

[1]John A. Paulson School of Engineering and Applied Sciences, Harvard University; Cambridge, Massachusetts, 02138, United States.

[2]Wyant College of Optical Sciences, The University of Arizona; Tucson, Arizona, 85721, United States.

[3]Air Force Research Laboratory, Wright-Patterson Air Force Base; Dayton, Ohio 45433, United States.

[4]Institute of Experimental Physics, Graz University of Technology, 8010 Graz, Austria.

†These authors contributed equally to this work.

*Corresponding author. Email: parkj@g.harvard.edu, capasso@seas.harvard.edu

**This PDF file includes:**

**Supplementary text:**

**Generation of metalens design file**
In this metalens design file generation, we place the nanopillars in cylindrical coordinates from azimuthal symmetry considerations[1]. The required fused silica nanopillar diameter $D(n)$ at each radial position $r_n$ is chosen to satisfy Eq. (1) as closely as possible. The edge-to-edge distance between the nanopillars is fixed at 250 nm. The $n^{th}$ nanopillar having diameter of $D(n)$ at each radial position $r_n$ is then placed evenly along the azimuthal direction so that the angular distance between nanopillars ($\Delta\theta_n$) satisfies the following equation:

$$\Delta\theta_n = \frac{2\pi}{\left\lfloor \frac{2\pi r_n}{D(n)+250\ nm} \right\rfloor},$$ (Eq. S1)

where $\lfloor x \rfloor$ is the floor function. This places nanopillars so that they have the same edge-to-edge distance equal to or slightly larger than 250 nm. This resulted in placement of 18,700,751,972 nanopillars for the entire 100 mm diameter metalens.

The created 100 mm diameter metalens design is then discretized into 25 sections, each having an area of $20 \times 20$ mm. Out of the 25 sections, only 7 sections (see Fig. S1) are selected and written to individual GDSII format CAD file using the GDSPY module on Python. We use 1 nm as the coordinate precision. Several techniques are used to reduce each CAD file size so that it can be later loaded to commercial photomask writers: (1) Each nanopillar with a specific diameter is only created once as a GDS cell, and other positions reference that first cell. (2) Each circle that represents a nanopillar is approximated to be an octagon to reduce the number of vertices. Nanopillars that are within 250 nm distance from the reticle boundaries are removed to avoid structure overlap between the sections. The total file size for all 7 reticle designs is 115.6 GB (19 GB, 20 GB, 20.3 GB, 20.1 GB, 16.5 GB, 3.2 GB, and 16.5 GB for reticles #1-#7, respectively), and all GDSII files are generated simultaneously using a 20-core workstation (two Intel® Xeon® E5-2690 v2 processors, 768 GB RAM) with a peak memory usage of 500 GB, taking less than 5 hours.

**Simulation of metalens performance**
The 100 mm metalens is numerically simulated by discretizing the 100 mm aperture into 100,000 annular rings with equal radial thickness of 500 nm, then discretizing each annular ring into 100 equal sections in the azimuthal direction, which produces 10 million annular sections. The radial discretization is chosen to resolve the highest spatial frequency zone at the edge of the metalens using at least four points. Each of the 10 million annular sections is associated with the complex scalar electric field that is transmitted through the nanopillar located at the geometric center of that annular section. The transmitted electric field through each nanopillar is obtained using the locally periodic assumption[2]. Each glass nanopillar on glass substrate geometry combination is individually simulated using the Rigorous Coupled Wave Analysis platform RETICOLO[3] on a periodic square lattice with a nominal edge-to-edge distance of 250 nm, 200 staircase discretization per circular arc quadrant, and 2601 Fourier plane waves. Empty regions without nanopillars are simulated under the same conditions with the nanopillar being replaced with a layer of air that has the same thickness as the nanopillar height. The fused silica refractive index as a function of wavelength is obtained from the dispersion equation published by Malitson[4] The transmitted electric field associated with each nanopillar is the zeroth diffraction order electric field for normal incidence illumination from the glass substrate. The area-weighted electric fields are propagated to the focal plane using a vectorial propagator[5]. The focal plane is sampled on a $201 \times 201$ square grid up to a maximum transverse diameter of 6 Airy disk diameters, where one Airy disk diameter is $1.22\lambda/NA$. The focusing efficiency is defined as the power flux through a circular diameter equal to 3 Airy disk diameters at the focal plane, divided by the incident power at the metalens. This focusing efficiency does not include the Fresnel reflection loss at the first air/glass interface on the back face of the metalens, which is expected to be around 4%. The 2D modulation transfer function (MTF) of the focal spot is computed by Fourier transformation of the intensity profile within 6 Airy disk diameters on the focal plane with no zero padding. The Strehl ratio of the focal spot is calculated by integrating the volume under the 2D MTF and dividing it by the volume under the diffraction limited 2D MTF.

| Publish Date (YYYY. MM.) | Diameter (λ) | Diameter (mm) | λ (nm) | Wavelength range | Fabrication Method | Material | Focal length (mm) | NA | f/# | Pol.-dep. (Y/N) | Focusing Eff. | Ref. |
|---|---|---|---|---|---|---|---|---|---|---|---|---|
| 1996. 02. | 1,580 | **1** (square) | 632.8 | Visible | EBL | Fused quartz | 20 | 0.025 | 20 | N | 53% | 6 |
| 1999. 05. | 233 | **0.2** (square) | 860 | NIR | EBL | TiO$_2$ | 0.4 | 20° off axis | | N | 80% | 7 |
| 2012. 08. | 581 | **0.9** | 1550 | NIR | EBL | Ag | 30, 60 | 0.015, 0.075 | 33.3, 66.7 | Y | 1% | 8 |
| 2013. 04. | 6 10 14 | **0.004 0.007 0.0094** | 676 | Visible | FIB | Au | 0.025 0.005 0.007 | 0.62 0.57 0.56 | 0.63 0.710 .74 | Y | 10% | 9 |
| 2014. 12. | 10 | **305** (square) | 30×10$^6$ (10 GHz) | GHz | Not reported | Metal | 300 | 0.58 | 0.70 | Y | 24.7% | 10 |
| 2016. 06. | 451 | **0.24** | 532 | Visible | EBL | TiO$_2$ | 0.09 | 0.80 | 0.38 | Y | 73% | 11 |
| 2017. 10. | 75 | **0.3** | 4000 | MIR | EBL | a-Si:H on MgF$_2$ | 0.05 | 0.95 | 0.17 | N | 78% | 12 |
| 2017. 10. | 104 | **10** | 96×10$^3$ (3.11 THz) | THz | Photolith. | SOI | 30 | 0.16 | 3.00 | N | 24% | 13 |
| 2018. 01. | 12,903 | **20** | 1550 | NIR | *i*-line stepper | a-Si | 50 | 0.20 | 2.50 | N | 91.8% | 14 |
| 2018. 02. | 839 | **0.6** | 715 | NIR | EBL | a-Si | 0.042 | 0.99 | 0.07 | N | 37% | 15 |
| 2018. 02. | 3,871 | **6** | 1550 | NIR | *i*-line stepper | a-Si | 50 | 0.06 | 8.33 | N | 62.5% | 16 |
| 2018. 02. | 571 -870 | **0.4** | 460 -700 | Visible | EBL | TiO$_2$ on SiO$_2$/Ag | 0.98 | 0.20 | 2.45 | N | 16-22.6% | 17 |
| 2018. 04. | 192 | **1** | 5200 | MIR | EBL | PbTe on CaF$_2$ | 0.5 | 0.71 | 0.50 | Y | 75% | 18 |
| 2018. 07. | 15,798 | **10** | 633 | Visible | EBL | SiN | -4 | 0.78 | -0.40 | N | *Not reported* | 19 |
| 2018. 11. | 30,303 37,594 42,283 | **20** | 660, 532, 473 | Visible | EBL | poly-Si | 12.9, 16, 18 | 0.61, 0.53, 0.49 | 0.65, 0.8, 0.9 | Y | 79% | 20 |

| Publish Date (YYYY. MM.) | Diameter (λ) | Diameter (mm) | λ (nm) | Wavelength range | Fabrication Method | Material | Focal length (mm) | NA | f/# | Pol.-dep. (Y/N) | Focusing Efficiency | Ref. |
|---|---|---|---|---|---|---|---|---|---|---|---|---|
| 2019. 08. | 2,000 1,200 750 | **6** (square) | 3000, 5000, 8000 | MIR | EBL | Ge | 2 | 0.83 | 0.33 | Y | 33% | 21 |
| 2019. 09. | 2,128 | **10** | 4700 | MIR | EBL | Au | 120 | 0.04 | 12.00 | Y | *Not reported* | 22 |
| 2019. 11. | 15,798 | **10** | 633 | Visible | KrF DUV | Fused silica | 50 | 0.10 | 5.00 | N | 45.6% | 1 |
| 2019. 11. | 5,639 | **3** | 532 | Visible | EBL | TiO$_2$ | 14.4 | 0.10 | 4.80 | N | *Not reported* | 23 |
| 2020. 01. | 2,128 | **2** | 940 | NIR | ArF immersion DUV | a-Si | 1.732 | 0.50 | 0.87 | N | 29.2% | 24 |
| 2020. 03. | 18,797 | **10** | 532 | Visible | EBL | Resist (ma-N) | 4 | 0.78 | 0.40 | Y | 80% | 25 |
| 2020. 05. | 19 -30 | **6.48** | 0.2×10$^6$- 0.3×10$^6$ (0.9 - 1.4 THz) | THz | Photolith. | Si | 15 | 0.21 | 2.31 | Y | 33.9% | 26 |
| 2020. 09. | 1,000 | **5.2** (square) | 5200 | MIR | EBL | PbTe on CaF$_2$ | 2.5 | 0.88 | 0.27 | N | 32-45% | 27 |
| 2021. 01. | 4,098 3,759 3,040 | **2** | 488, 532, 658 | Visible | EBL | TiO$_2$ | 3.2, 1 | 0.3 0.7 | 0.5 1.55 | Y | *Not reported* | 28 |
| 2021. 02. | 288 | **1.5** (square) | 5200 | MIR | EBL | GSST | 1.5, 2 | 0.58, 0.47 | 0.71, 0.94 | Y | 23.7%, 21.6% | 29 |
| 2021. 04. | 581 | **0.9** | 1550 | NIR | NIL | Si | 10 | 0.04 | 11.11 | Y | 26% | 30 |
| 2021. 08. | 7,273 | **4** | 550 | Visible | NIL | TiO$_2$ | 9.8 | 0.20 | 2.45 | N | 43% | 31 |
| 2021. 10. | 1,290 | **2** | 1550 | NIR | EBL | a-Si | 4 | 0.24 | 2.00 | N | 42.7% | 32 |
| 2022. 01. | 3,160 | **2** | 633 | Visible | EBL | sc-Si | 2 | 0.45 | 1.00 | N | 64% | 33 |
| 2022. 04. | 4,887 | **2.6** | 532 | Visible | EBL | GaN | 5.03 | 0.25 | 1.93 | N | *Not reported* | 34 |

| Publish Date (YYYY. MM.) | Diameter (λ) | Diameter (mm) | λ (nm) | Wavelength range | Fabrication Method | Material | Focal length (mm) | NA | f/# | Pol.-dep. (Y/N) | Focusing Eff. | Ref. |
|---|---|---|---|---|---|---|---|---|---|---|---|---|
| 2022. 05. | 20,492 18,797 15,198 | **10** | 488, 532, 658 | Visible | EBL | TiO₂ | 16 | 0.30 | 1.60 | N | 15% | 35 |
| 2022. 05. | 55,172 | **80** | 1450 | NIR | KrF DUV | a-Si | 260 | 0.15 | 3.25 | N | 71% | 36, 37 |
| 2022. 05. | 4,717 | **50** | 10600 | LWIR | Stepper photolith. | Si | 34 | 0.59 | 0.68 | N | *Not reported* | 38 |
| 2022. 08 | 8,000 | **80** | 10000 | LWIR | Contact photolith. | a-Si | 80 | 0.45 | 1.00 | N | *Not reported* | 39 |
| 2023. 01 | 4,887 | **2.6** | 532 | Visible | EBL | GaN | 10 | 0.13 | 3.85 | N | 74% | 40 |
| 2023. 03 | 22,222 18,797 15,748 | **10** | 450, 532, 635 | Visible | NIL, ArF immersion DUV | TiO₂ coated resin | 24.5 | 0.20 | 2.45 | Y | 40.9%, 55.6%, 44.6% | 41 |
| **This work** | 157,978 | **100** | 633 | Visible | KrF DUV | Fused silica | 150 | 0.32 | 1.50 | N | 40.4% | |

**Table. S1. List of notable metalens' diameters and their parameters.**

Progression in time of notable metalens diameters with respect to design wavelength (λ and mm), fabrication methods, constituent materials, polarization dependence (pol.-dep.), and their optical parameters (focal length, numerical aperture (NA), f-number (*f*/#), and focusing efficiency. Acronyms in the fabrication method column are as follows:

- EBL: Electron beam lithography
- NIL: Nano-imprint lithography
- FIB: Focused-ion beam lithography
- DUV: Deep-ultraviolet lithography

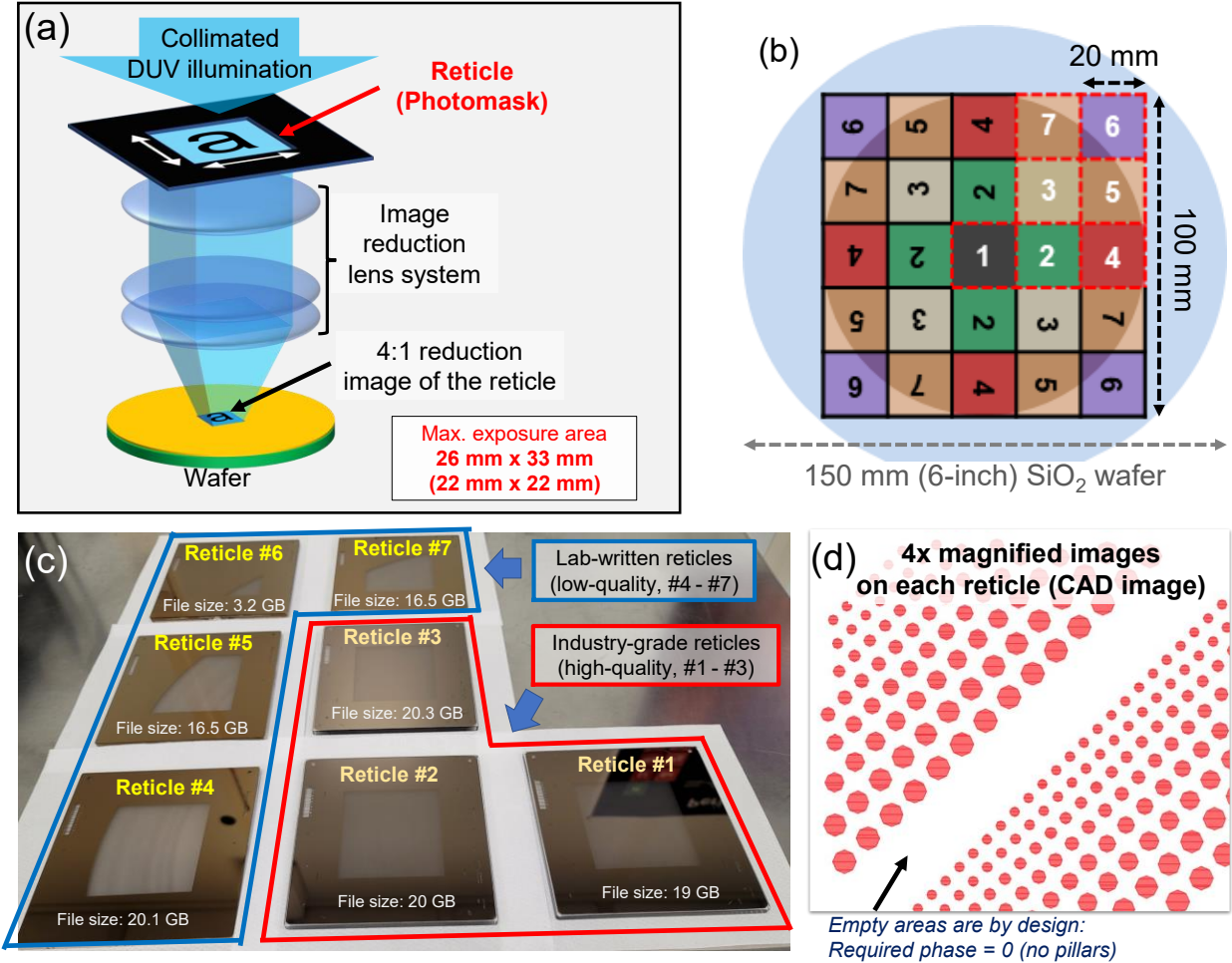

**Fig. S1. Fabrication strategy and prepared reticles.**

(a) An abbreviated schematic of a DUV projection lithography system. A DUV beam ($\lambda$=*248 nm* for KrF excimer laser source) is incident on a reticle (photomask) which has the desired patterns written in chrome. The transmitted light is then passed through an array of image reduction lenses so that the image on the reticle is demagnified and projected onto the surface of a target wafer. Depending on the lithography tool, the demagnification ratio is typically 4:1 or 5:1. This allows the resolution of features smaller than the source wavelength, while keeping the reticles pristine. These image reduction systems, however, limit the maximum exposure area with guaranteed image resolution only on the order of *30 mm* square area on the wafer. In this article, we use an ASML PAS 5500/300C DUV stepper having $\lambda$=*248 nm*, demagnification ratio of 4:1, and maximum exposure area of $22 \times 22\ mm$ on a wafer. (b) To realize a 100 mm diameter metalens using the DUV system, we discretize the lens into 25 different sections, each having a $20 \times 20\ mm$ area. We intentionally put reticle #1 at the center of the metalens pattern to avoid stitching errors on the optic axis. As the metalens is azimuthally symmetric, we identify and use only the 7 unique sections from a quadrant to make the whole metalens. As we target a 100 mm (4-in.) diameter metalens, we choose the wafer to be larger than the metalens diameter (150 mm, 6-in.) which allows space for alignment markers and easier handling. (c) Fabricated reticles for the 100 mm metalens. Reticles #1-#3 (inner $60 \times 60$ mm area) are fabricated by industry-grade photomask manufacturer (Photronics Inc.) and reticles #4-#7 are fabricated in-house using a Heidelberg DWL2000 laser mask writer. Reticles #4-#7 have poorer quality (unresolved small patterns, incorrect feature sizes) compared to reticles #1-#3 due to imperfection in the optimization of in-house mask fabrication processes. The prepared CAD file size (GDSII format) is marked on each reticle. (d) Image of the reticle patterns on a top left edge of the reticle #1. Filled areas (red) represent cleared areas on the reticle, while the unfilled (white) area is opaque with chrome mask. Empty areas along the azimuth are part of the design that contributes to the wavefront. The circular patterns are approximated as octagons to reduce the number of vertices, and therefore reduce the overall CAD file size. The patterns are 4X magnified on the reticle. Structures that overlap with the reticle boundaries are intentionally omitted to avoid multiple exposure between the reticles.

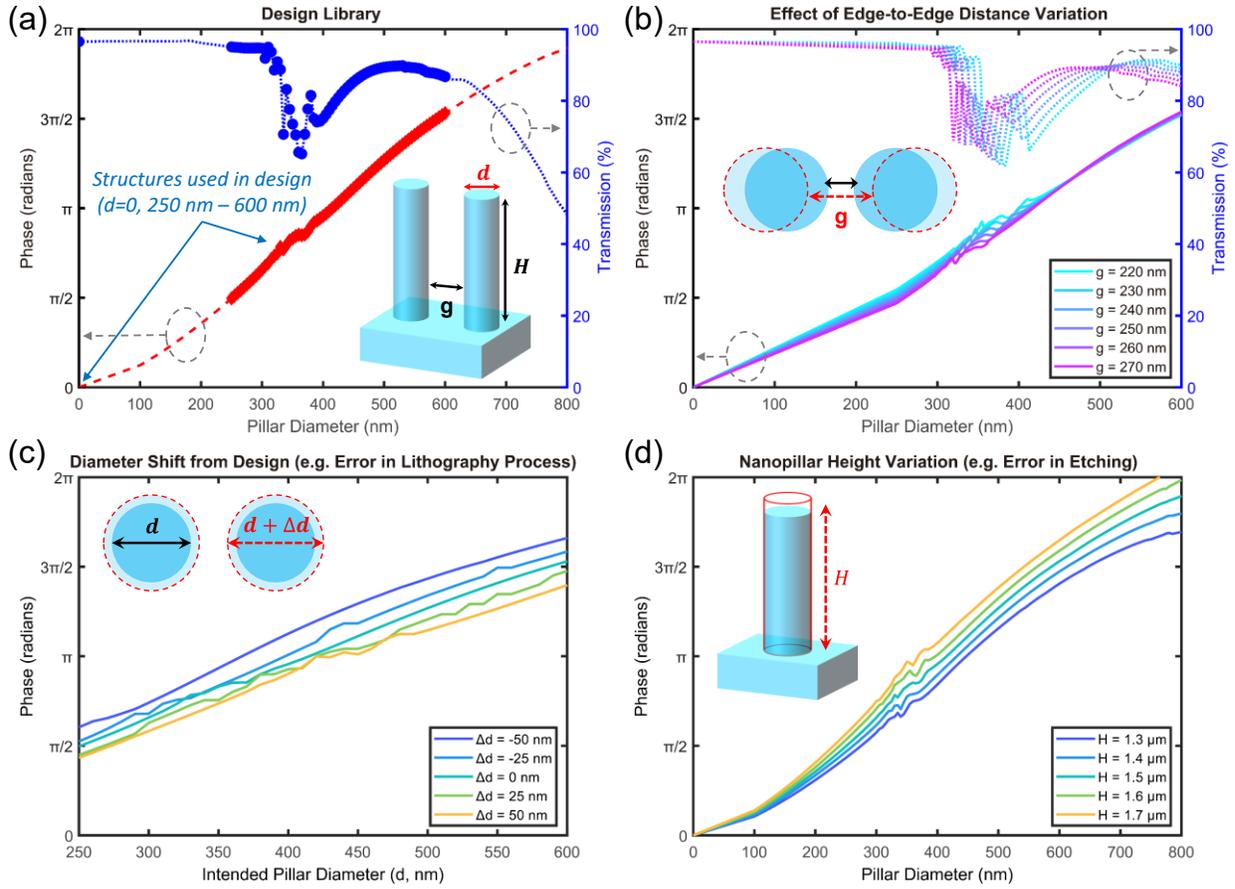

**Fig. S2. Simulated library of fused silica nanopillars.**
(a) Library of nanopillars used in the metalens design. Nanopillar height H=*1.5 μm* and edge-to-edge gap g=250 nm is used for while sweeping the diameters from 0 to 800 nm. Only the structures with diameters between 250 nm and 600 nm, and 0 nm (empty area) are selected to be included in the metalens design. (b) Effect of edge-to-edge gap variance on the nanopillar's transmitted phase and intensity. (c) Effect of diameter shift while maintaining the same center-to-center distance to model the size error which could occur during fabrication. (d) Effect of overall nanopillar's height shift on the transmitted phase to model the height error which could occur during fabrication.

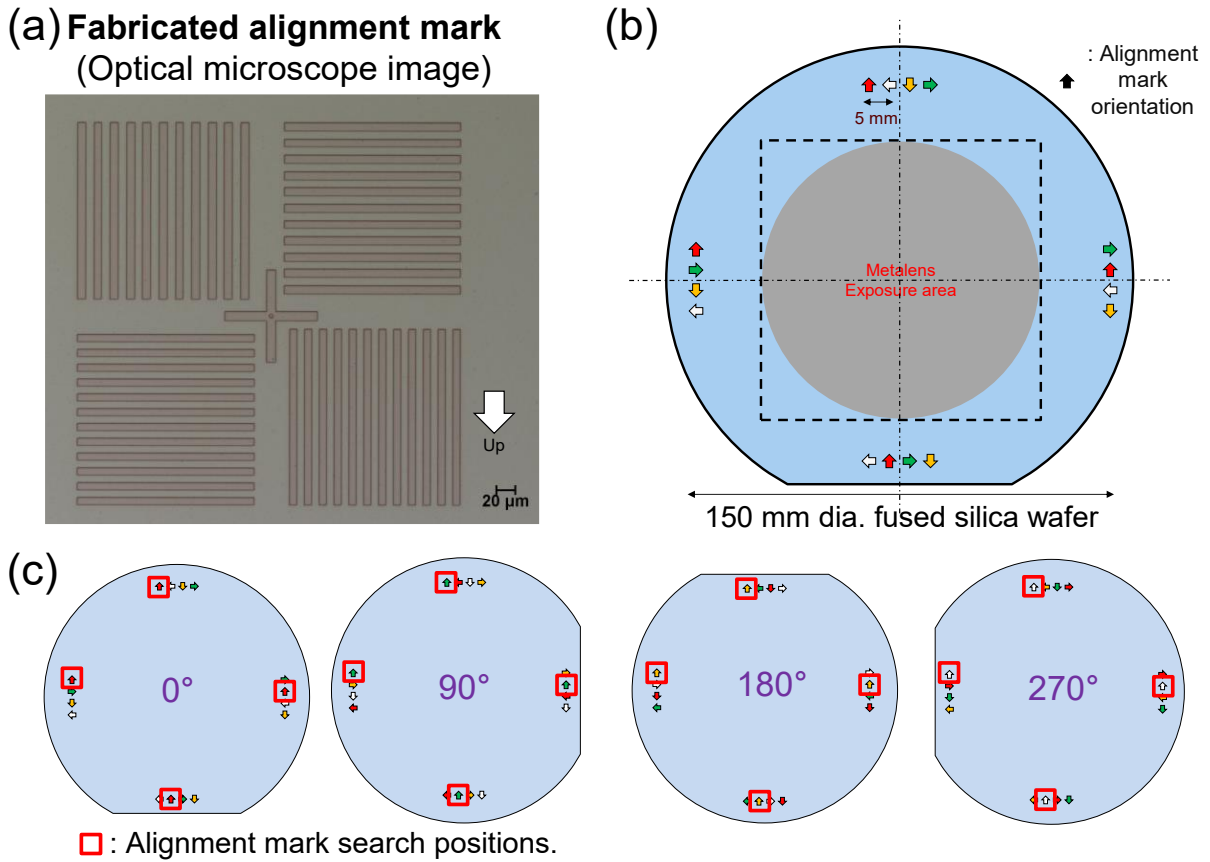

**Fig. S3. Rotation and alignment strategy.**

(a) Optical microscope image of the alignment marks for the ASML PAS5500/300C DUV stepper. After coating DUV-42P ARC (Brewer Science, anti-reflection coating) and UV210 resist (Shipley, positive DUV resist), the alignment patterns are exposed on the 150 nm thick aluminum film on a 150 mm diameter fused silica wafer using the DUV stepper system (Fig. 1(a)). After post-exposure bake and development of the alignment marks, the patterns are transferred to the Al film using wet etching. The alignment marks are etched to 120 nm in depth to satisfy the phase-contrast detection system in the DUV stepper. (b) Schematic of alignment mark placements. An arrow indicates a mark's position, and the orientation of the mark at each position is indicated by the direction of the arrow. The marks are placed outside the metalens exposure area and are used as global alignment keys during the projection lithography processes for each reticle. The marks are placed in a way that the DUV tool can search the same location for the four alignment marks at each wafer rotation of 0°, 90°, 180°, and 270°, respectively, for the convenience in exposure schedule programming.

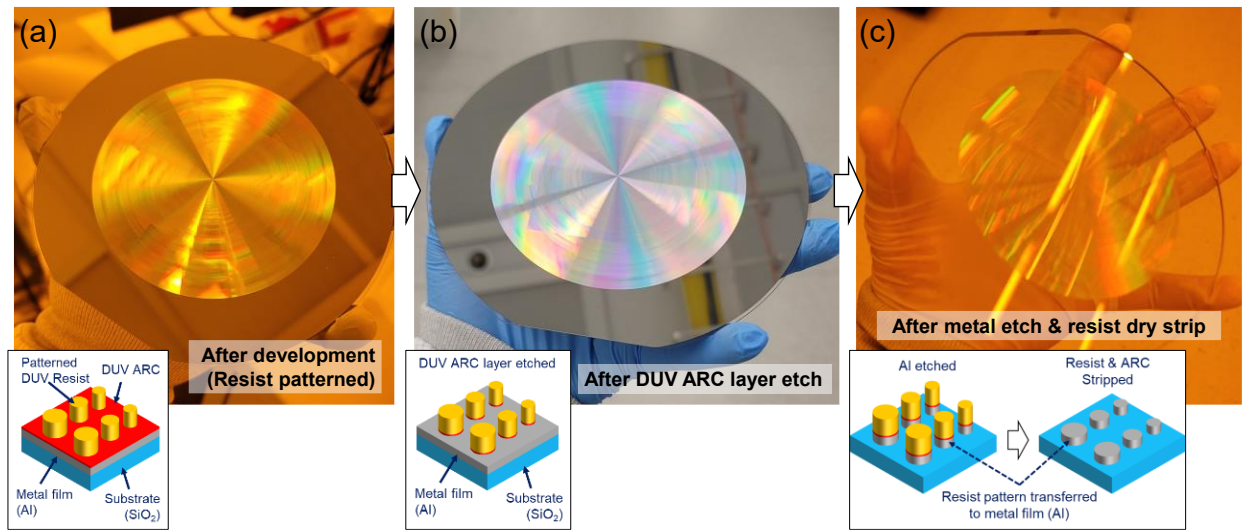

**Fig. S4. Photographs of metalens wafer in between fabrication steps.**
(a) Photograph of the developed photoresist pattern of the 100 mm metalens. The pattern is formed on the aluminum (Al) film on a 150 mm (6-inch) fused silica wafer, coated with DUV antireflective film (DUV ARC). (b) Using $O_2$/Ar plasma, the DUV ARC layer is etched so that the metalens pattern is transferred to the ARC layer and expose the Al film below. (c) The 100 mm metalens pattern is then transferred to the Al film using $Cl_2$ reactive ion etch. The Al on the edge of the wafer is not etched due to the wafer clamping mechanism of the plasma etcher. The residual resist and ARC materials are stripped using downstream plasma ashing.

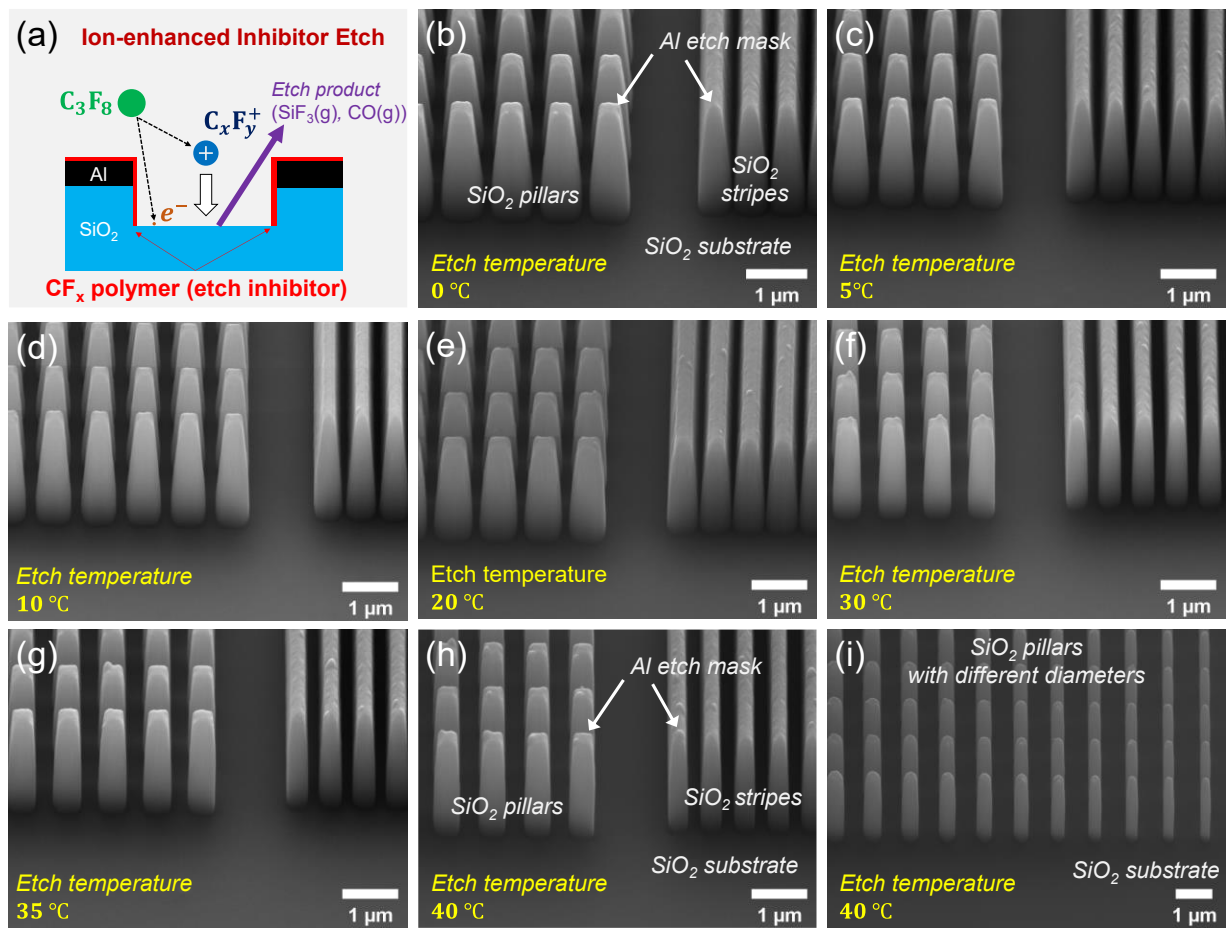

**Fig. S5. Optimization of fused silica vertical etching process.**

(a) Schematic of ion-enhanced inhibitor etch process. $C_3F_8$ gas is introduced into the etch chamber and is ionized by RF drive. Light-weight, free electrons are first attracted toward the electrically isolated wafer surface and charge the surface to negative polarity as the electron population is built up. The ionized gas simultaneously forms fluorocarbon polymer on the exposed surface, creating a chemical etch inhibiting film, while the positively charged fluorocarbon ions are accelerated toward the negatively charged surface. The accelerated ions directionally bombard the surface, breaking and removing the fluorocarbon polymer film that faces the surface normal of the substrate. The fluoride radicals present in the plasma diffuse onto the surface and chemically etch the exposed $SiO_2$, creating volatile $SiF_3$ and CO as etch products. The polymers deposited on the sidewalls inhibit the lateral etch of $SiO_2$. The vertical etch of $SiO_2$ is achieved by balancing the rate of fluorocarbon deposition and the reactive ion etch. Here, we use substrate temperature as the rate control parameter. (b)-(h) Tilted SEM images of the etched $SiO_2$ pillars and trenches at different etch temperatures. The fixed process temperature for each sample ranges from 0°C to 40°C. As the temperature rises, the polymer deposition speed is slowed and the reactive ion etching is accelerated, that near-vertical sidewalls are reached at 40°C. The samples in the SEM images still have residual Al etch mask on the top of each structure, which accounts for the rough top surface. (i) Tilted SEM image of the vertical-sidewall sidewall etched $SiO_2$ nanopillars with different diameters. The height of the nanopillars is 1.5 $\mu m$. The smooth area at the bottom of the pillars is an etched surface.

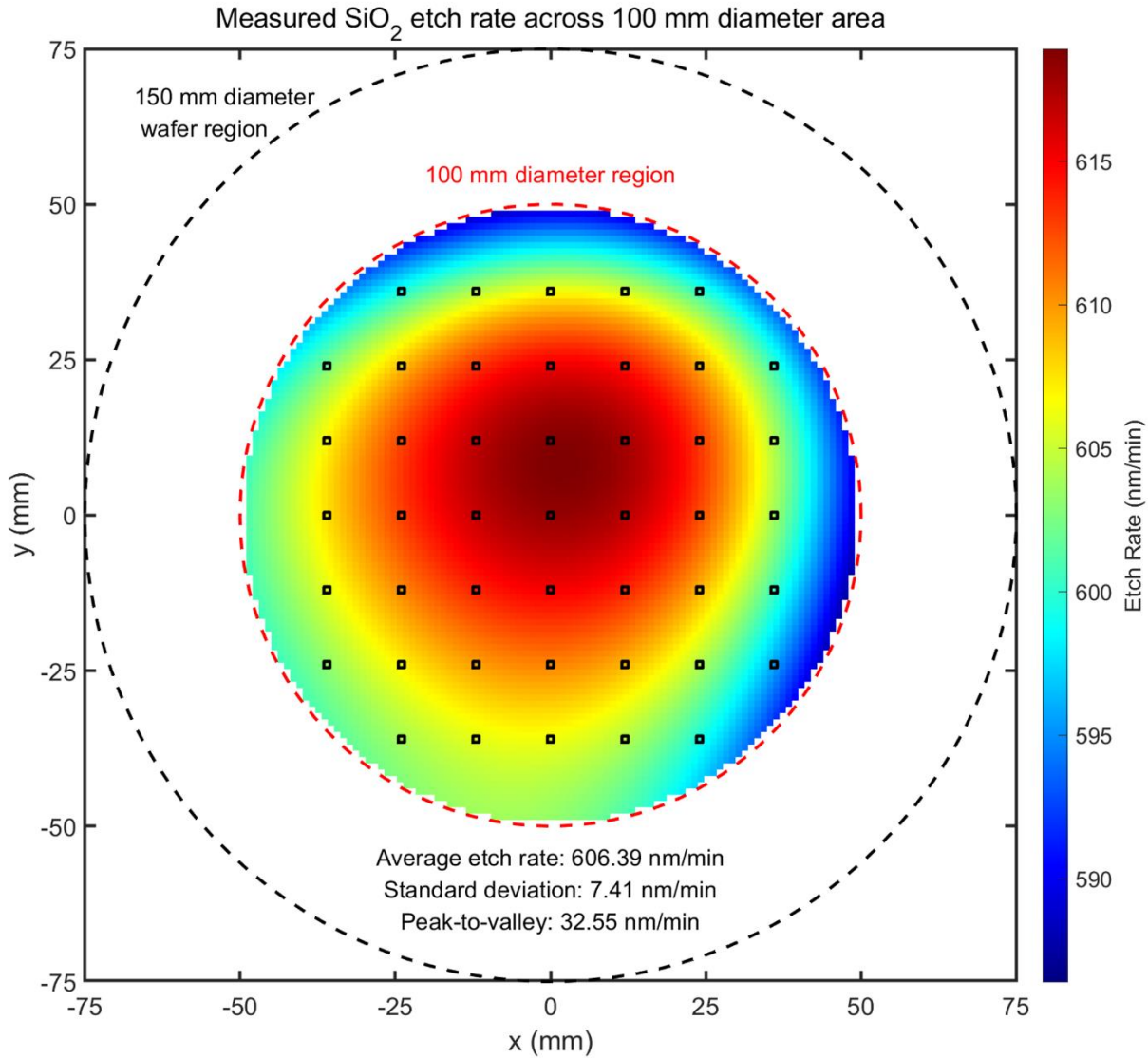

**Fig. S6. Measured etch rate uniformity across the 100 mm diameter region.**
The SiO₂ etch rate is measured across 100 mm diameter region with stylus profilometry at 45 different positions. The positions of the acquired data are marked in small green squares. With the measurement data, the etch rate across the region of interest is obtained by fitting to a cubic polynomial:

$$f(x, y) = a_0 + a_1 x + a_2 y + a_3 xy + a_4 x^2 + a_5 y^2 + a_6 x^3 + a_7 x^2 y + a_8 xy^2 + a_6 y^3.$$

The measured average etch rate across the 100 mm diameter region is 606.4 nm/min, with standard deviation of 7.41 nm/min. The etch uniformity could be further improved by having better control over the temperature distribution and plasma uniformity in the etch chamber.

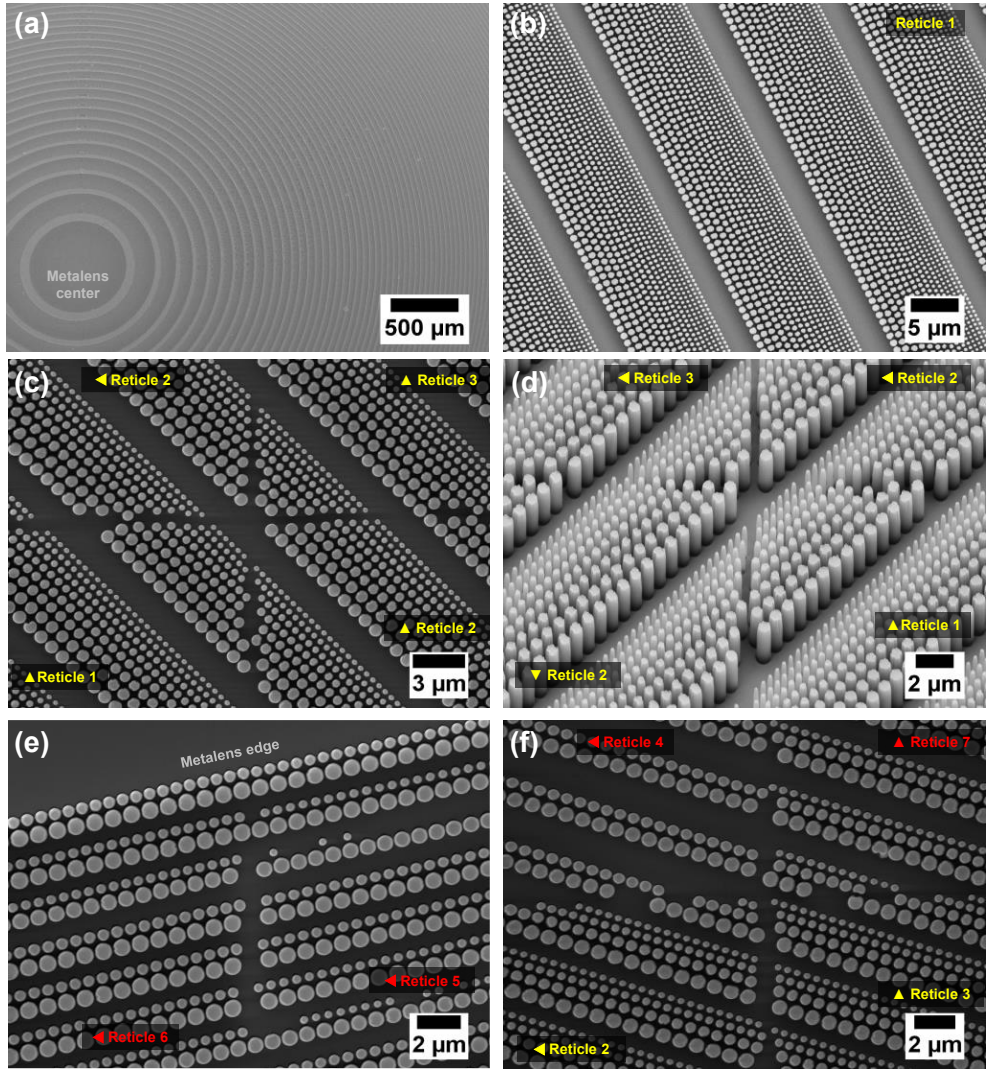

**Fig. S7. SEM images of the fabricated 100 mm diameter all-glass metalens.**
(a) Top-view SEM image of the metalens center area, corresponding to Reticle #1. (b) Top-view SEM image of the metalens. (c) Top-view SEM image of the metalens where 4 reticles meet; 3 reticles are oriented north, while reticle #2 (top left) is oriented toward west. The missing pillars between the reticles are intentional; they were omitted to avoid possible multiple exposure at the boundary of the sections. (d) Tilted SEM image of metalens where 4 industry-grade reticles meet; reticle #1 is oriented north, while the top two reticles (#2 and #3) are facing west, and lower left corner reticle (#2) is facing south. Note that the vertical sidewalls are consistent for different pillar diameters. (e) Top-view SEM image of the metalens at its edge, between the boundaries of home-made reticles #6 and #5. The missing small pillars are from fabrication error of the reticles, where small pillar structures were not resolved. (f) Top-view SEM image of the metalens where home-made (#4, #7) and industry-grade reticles (#2, #3) meet. Comparing with reticles 2 and 3, reticles 4 and 7 shows slight misalignment, missing small pillars, and incorrect pillar size, originating from the lower-resolution mask writing process developed in-house.

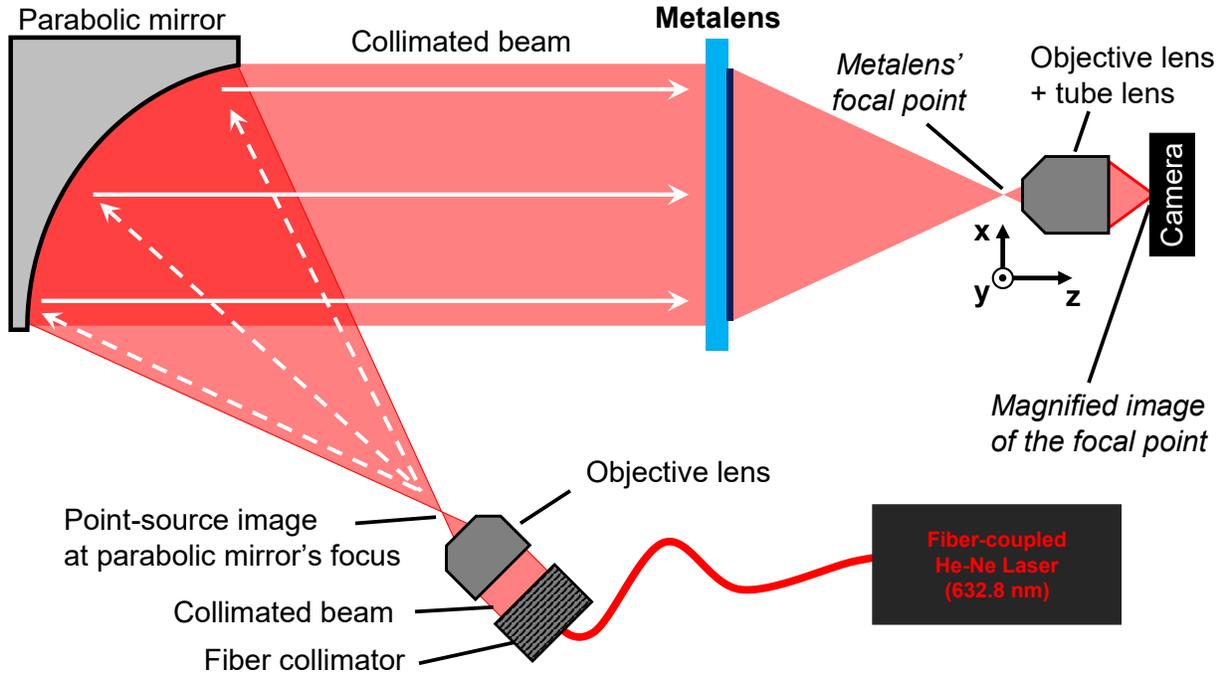

**Fig. S8. Schematic of point-spread-function measurement setup.**
A fiber-coupled He-Ne laser ($\lambda = 632.8\ nm$, N-LHR-121, *Newport Corporation*) is used as an illumination source. After passing the laser beam through the fiber collimator (F220APC-633, *Thorlabs*), the collimated beam is then sent through an objective lens (EA achro 20X/0.40, W.D. 0.8 mm, *Motic*) from its back side. The focal point of the objective lens is used as the point-source image, which is placed at the focus of a 45° off-axis parabolic mirror (#35-629, *Edmund Optics*). The parabolic mirror then creates a collimated beam with diameter of 101 mm, which is incident on the 100 mm diameter metalens. The image of the focus is 40X magnified and recorded by the microscope camera of Optikos MTF measurement tool (LensCheck, *Optikos Corp.*). The microscope camera is stepped along z-axis to acquire the point-spread-function along the optic axis. The focal plane of the metalens is determined by the plane of maximum on-axis intensity.

For an accurate PSF measurement, a high-quality collimated light incident on a metalens at a normal angle is important. This is relatively easier to achieve for few millimeter aperture optics, where quality optomechanical components are readily available, but not for a hundred-millimeter or larger scale optics: Not only is the creation of large-diameter collimated beam with near-planewave phase front difficult and costly, the alignment of large optics with respect to the collimated beam also requires extensive trial and error. To ensure the validity of the PSF characterization, therefore, a different measurement technique is needed to corroborate the PSF analysis for large-diameter optics, in general. A more commonly used and widely accepted characterization method for large optics is interferometric analysis.

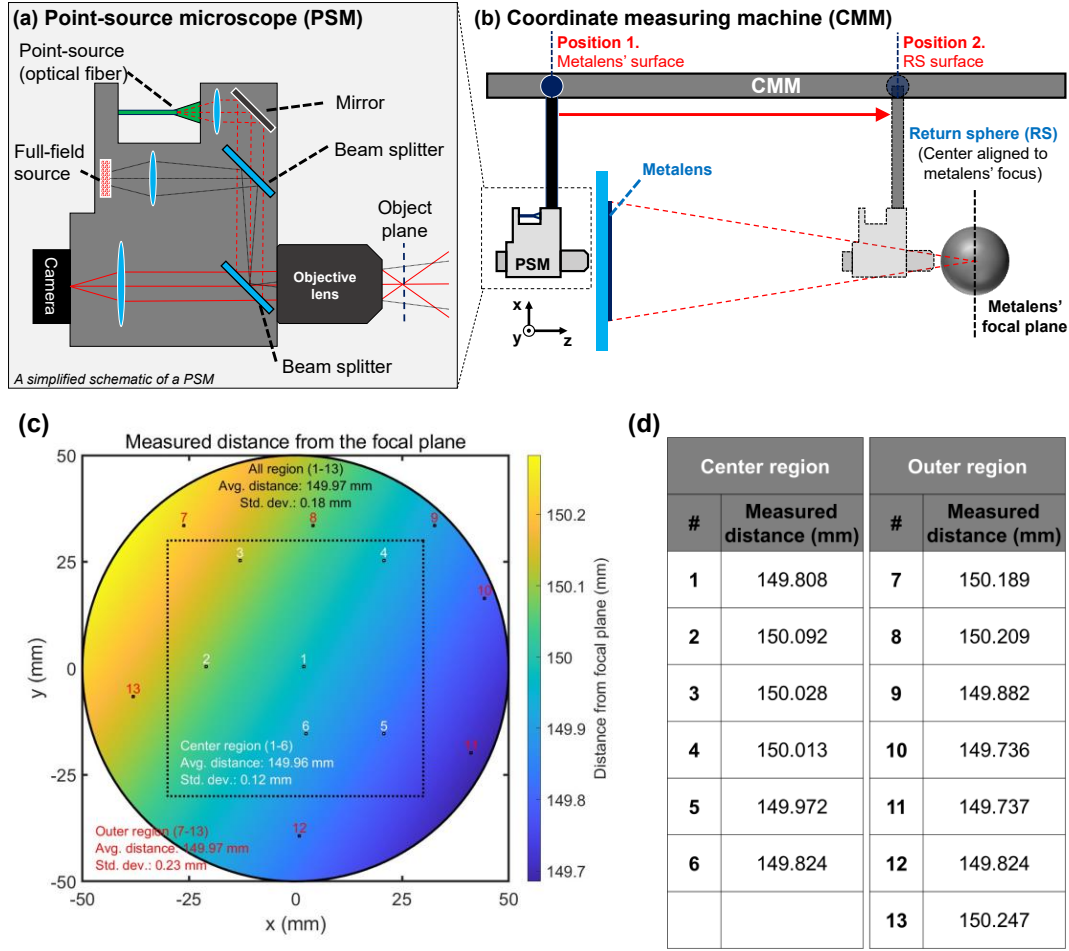

**Fig. S9. Focal length measurement using point-source microscope (PSM) and coordinate measurement machine (CMM).**
(a) A schematic diagram of a point-source microscope (PSM, *Optical Perspectives Group LLC*). A point source image is projected onto the objective lens' object plane, which is used as a reference to ensure the PSM is correctly focused on the surface of interest or at the center of a sphere. Detailed descriptions and discussions can be found in Ref. 42. (b) Schematic of the focal length measurement apparatus. A PSM is securely mounted on a coordinate measurement machine (CMM, *Tesa Micro-hite 3D, Tesa Technology*), which is first focused on the rear surface of the metalens. After acquiring the x,y,z positions of the PSM at various metalens surface positions, the PSM is then moved to focus at the center of the return sphere (RS), whose center is aligned to the metalens' focus using a Fizeau interferometer. The known thickness of the metalens substrate is subtracted out. (c) Various points of measurements at the surface of the metalens. The positional data are taken at 13 different locations and are labeled with numbers: 1-6 for the inner section of the metalens, and 7-13 for the outer section of the metalens. The measured distance from the focal plane (*i.e.*, metalens' surface to the return sphere's center of curvature) is 149.97 mm with standard deviation of 0.18 mm, which is in good agreement with the designed focal length of 150 mm. (d) List of measured axial (i.e., along the metalens' surface normal) distance values from the return sphere's center of curvature at different positions of the metalens.

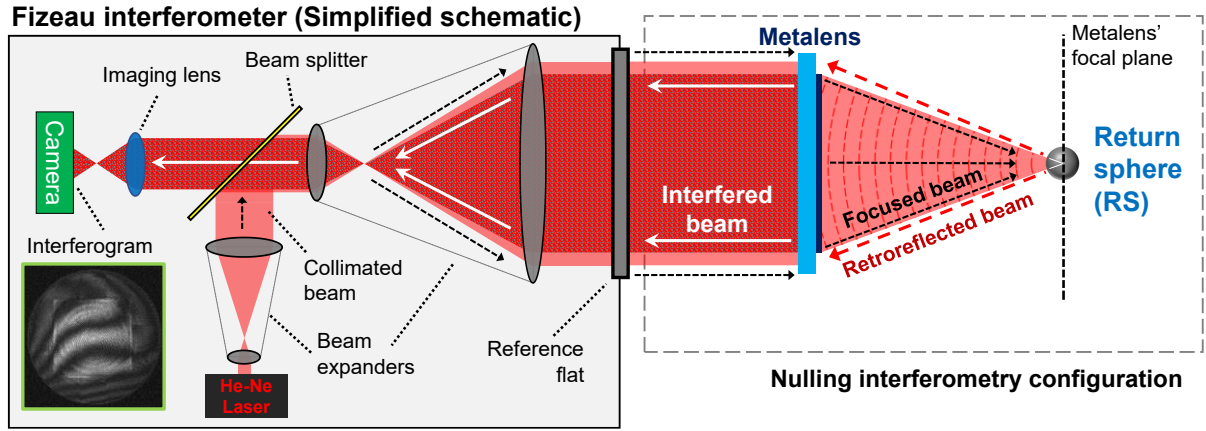

**Fig. S10. Schematic of full-aperture interferometry measurement setup.**

A laser Fizeau interferometer (Zygo Verifire™ ATZ, *Zygo Corporation*) was used in nulling interferometry double-pass configuration to obtain the full-aperture wavefront measurement. We note that the image depicted on the left is not an exact configuration of the used interferometer but is only a simplified schematic diagram to illustrate the measurement concept. For better understanding, please see Ref. 42. The metalens is illuminated by a collimated beam ($\lambda = 632.8\ nm$) from a Fizeau interferometer, creating the metalens' focus at its focal plane. A return sphere (RS) with a known diameter is placed at the center of the metalens' focus, having the focal plane coincide with the center of the return sphere. The focused beam by the metalens propagates toward the return sphere and is retroreflected from the RS surface, which is re-incident on the metalens. The doubly transmitted (*i.e.*, double-pass) wavefront creates twice of the wavefront aberration function (WAF) deviating from an ideal spherical wavefront; if the wavefront from the metalens is perfectly spherical, the reflected wavefront from the return sphere would also be perfectly spherical, resulting in a null interferogram. The interfered irradiance image is then acquired with a camera. Multiple phase-shifted fringe patterns are used to calculate the phase map. The phase map is unwrapped across the full-aperture to provide the wavefront error map. After removing artifacts from misalignment (*i.e.*, piston, tip, tilt, and power) of the metrology configuration (*i.e.*, interferometer, metalens, and the return sphere) we obtain the metalens' wavefront error.

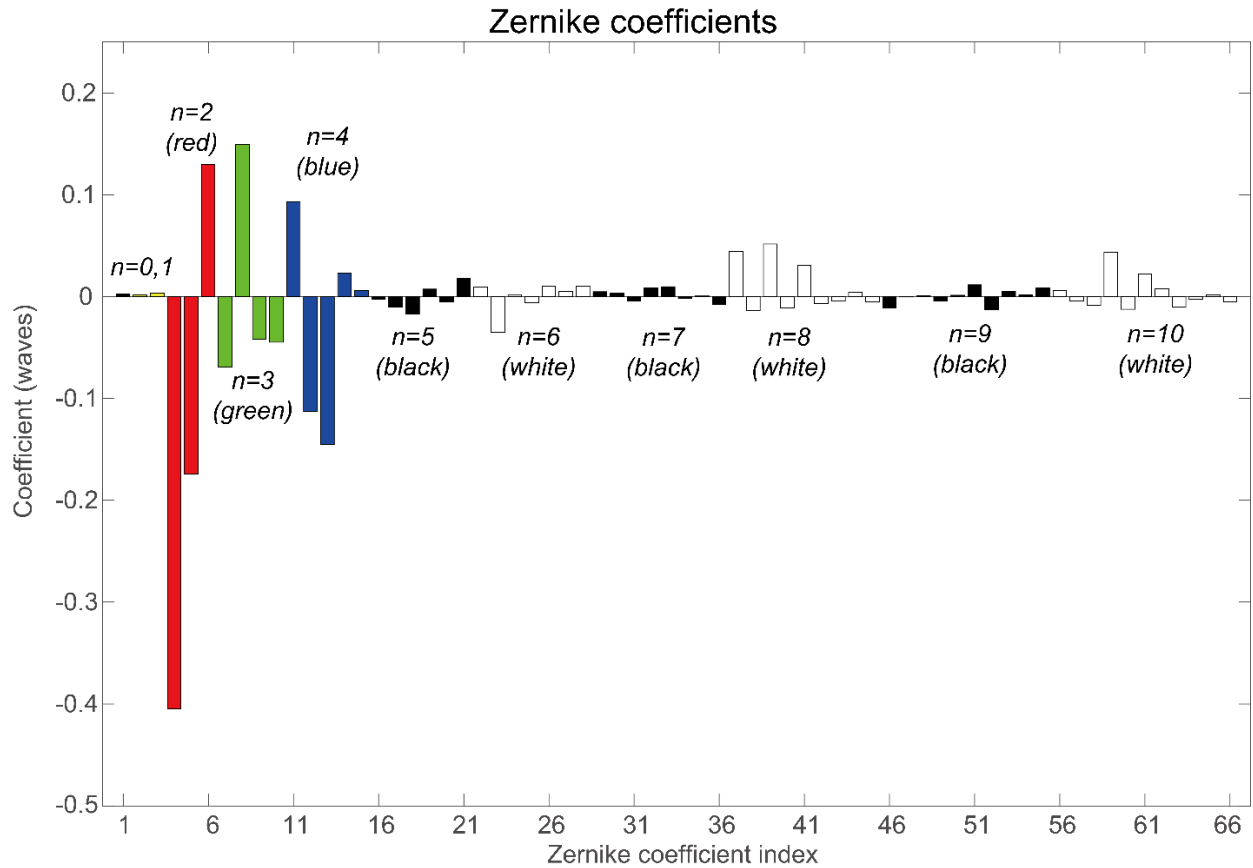

**Fig. S11. Fitted coefficients of Zernike polynomials of the measured WAF up to 66th order (10th degree)**

The bar plots correspond to the fitted WAF coefficient values in Table S2. The bars are colored by the radial degree ($n$) of Zernike polynomial for visual discrimination. Major contribution to aberration comes from the 2nd (colored red, index no. 4-6) and the 4th (colored blue, index no. 11-15) radial degrees. The large contribution from the horizontal coma (colored green, 3rd degree, index no. 8) is likely from the field-dependent aberration due to the misalignment of the metalens in the interferometry setup.

| Index | Degree (n) | Order (m) | Coefficient (waves) | Common name |
|---|---|---|---|---|
| 1 | 0 | 0 | 0.002 | *Piston* |
| 2 | 1 | 1 | 0.002 | *Horizontal (x) tilt* |
| 3 | | -1 | 0.003 | *Vertical (y) tilt* |
| 4 | 2 | **2** | **-0.405** | ***Vertical astigmatism*** |
| 5 | | **0** | **-0.174** | ***Defocus*** |
| 6 | | **-2** | **0.130** | ***Oblique astigmatism*** |
| 7 | 3 | 3 | -0.069 | *Oblique trefoil* |
| 8 | | **1** | **0.149** | ***Horizontal coma*** |
| 9 | | -1 | -0.041 | *Vertical coma* |
| 10 | | -3 | -0.045 | *Vertical trefoil* |
| 11 | 4 | **4** | **0.093** | ***Vertical quadrafoil*** |
| 12 | | **2** | **-0.113** | ***Vertical secondary astigmatism*** |
| 13 | | **0** | **-0.145** | ***Primary spherical*** |
| 14 | | -2 | 0.023 | *Oblique secondary astigmatism* |
| 15 | | -4 | 0.006 | *Oblique quadrafoil* |
| 16 | 5 | 5 | -0.003 | |
| 17 | | 3 | -0.010 | |
| 18 | | 1 | -0.017 | *Horizontal secondary coma* |
| 19 | | -1 | 0.007 | *Vertical secondary coma* |
| 20 | | -3 | -0.005 | |
| 21 | | -5 | 0.018 | |
| 22 | 6 | 6 | 0.009 | |
| 23 | | 4 | -0.035 | |
| 24 | | 2 | 0.002 | |
| 25 | | 0 | -0.006 | *Secondary Spherical* |
| 26 | | -2 | 0.010 | |
| 27 | | -4 | 0.005 | |
| 28 | | -6 | 0.010 | |
| 29 | 7 | 7 | 0.005 | |
| 30 | | 5 | 0.004 | |
| 31 | | 3 | -0.004 | |
| 32 | | 1 | 0.009 | |
| 33 | | -1 | 0.010 | |

| Index | Degree (n) | Order (m) | Coefficient (waves) |
|---|---|---|---|
| 34 | 7 | -3 | -0.002 |
| 35 | | -5 | 0.001 |
| 36 | | -7 | -0.008 |
| 37 | 8 | 8 | **0.044** |
| 38 | | 6 | **-0.013** |
| 39 | | 4 | **0.052** |
| 40 | | 2 | **-0.011** |
| 41 | | 0 | **0.030** |
| 42 | | -2 | **-0.007** |
| 43 | | -4 | **-0.005** |
| 44 | | -6 | **0.004** |
| 45 | | -8 | **-0.005** |
| 46 | 9 | 9 | **-0.011** |
| 47 | | 7 | **0.000** |
| 48 | | 5 | **0.001** |
| 49 | | 3 | **-0.004** |
| 50 | | 1 | **0.001** |
| 51 | | -1 | **0.012** |
| 52 | | -3 | **-0.013** |
| 53 | | -5 | **0.005** |
| 54 | | -7 | **0.002** |
| 55 | | -9 | **0.008** |
| 56 | 10 | 10 | **0.006** |
| 57 | | 8 | **-0.005** |
| 58 | | 6 | **-0.008** |
| 59 | | 4 | **0.044** |
| 60 | | 2 | **-0.012** |
| 61 | | 0 | **0.022** |
| 62 | | -2 | **0.008** |
| 63 | | -4 | **-0.010** |
| 64 | | -6 | **-0.003** |
| 65 | | -8 | **0.002** |
| 66 | | -10 | **-0.005** |

**Table. S2. Fitted coefficients of Zernike polynomials of the measured WAF up to 66th order (10th degree)**

Zernike polynomial basis are expressed as:

$$\begin{cases} Z_n^m(\rho, \varphi) = R_n^m(\rho)\cos(m\varphi) \ (\text{for } m \geq 0) \\ Z_n^m(\rho, \varphi) = R_n^m(\rho)\sin(m\varphi) \ (\text{for } m < 0) \end{cases} \tag{Eq. S2}$$

Where the radial polynomial $R_n^m(\rho)$ is,

$$R_n^m(\rho) = \sum_{s=0}^{(n-m)/2}(-1)^s \frac{(n-s)!}{s!\left(\frac{n+m}{2}-s\right)!\left(\frac{n-m}{2}-s\right)!}\rho^{n-2s}. \tag{Eq. S3}$$

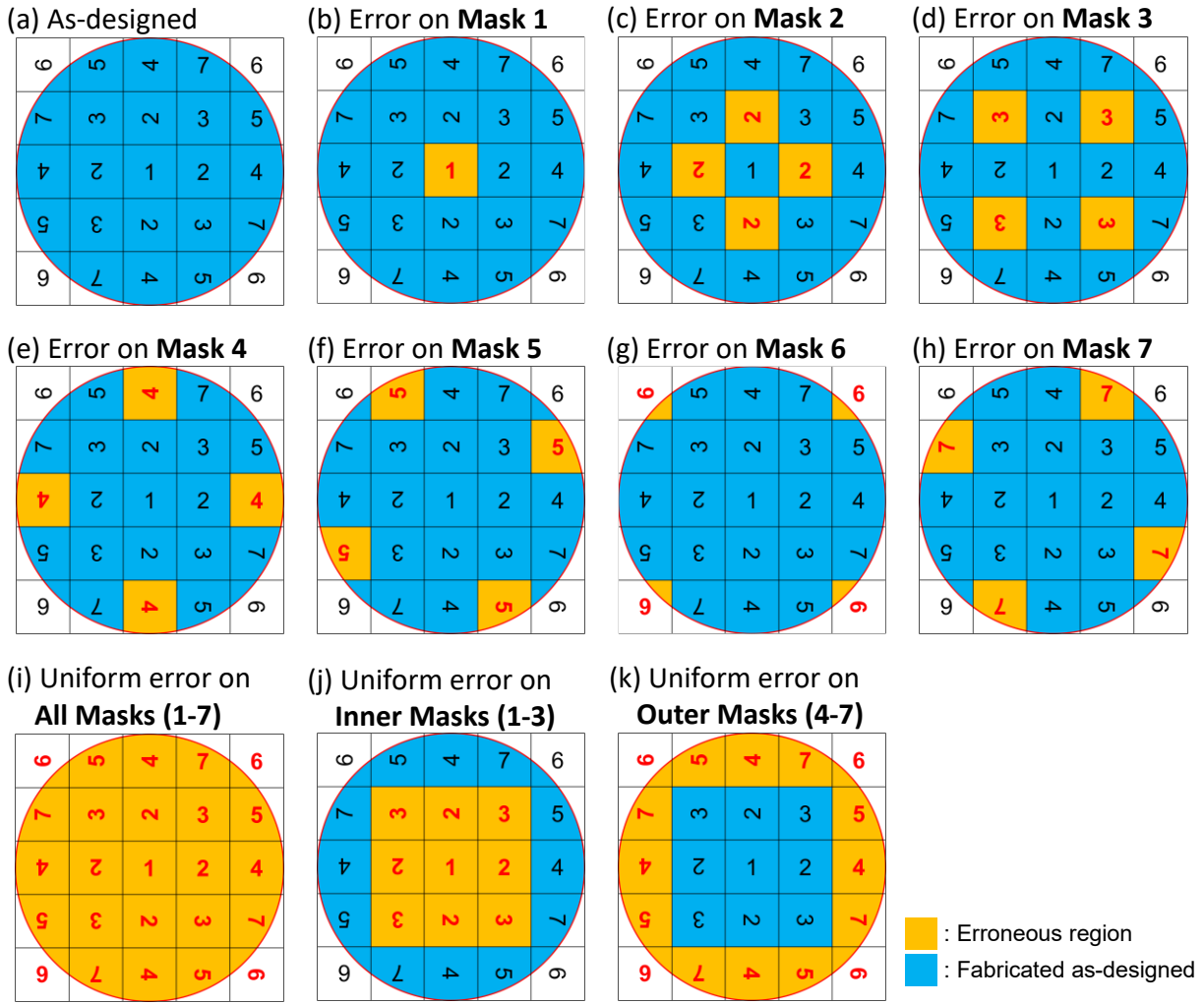

**Fig. S12. Schematic of considered fabrication error scenarios.**
The considered error scenarios in the simulation-based study are as follows: (1) Small pillars are not resolved (not meeting critical dimension criterion), (2) shift in nanopillar diameters and (3) error in nanopillar height (due to error during etching). (a) Distribution of reticle patterns of an as-designed metalens, having no fabrication error. Reticle (Mask) 1 is placed at the center and Mask 2-7 are distributed at each quadrant rotated by 0°, 90°, 180° and 270°, respectively. (b)-(h) Distribution of an erroneous mask across the metalens, when Mask 1 to Mask 7, respectively, is wrongly fabricated. (i) Distribution of erroneous masks when all 7 are erroneous. Distribution of erroneous masks when (j) Masks 1-3 (inner sections) and (k) Masks 4-7 (outer sections) have error. The dominating fabrication error of the measured metalens in the manuscript corresponds to the case of (k) with error scenario (1).

| Non-ideality | Fields affected | Error sources | Focusing efficiency | Strehl ratio |
|---|---|---|---|---|
| **Smaller pillars missing** | All | Poor reticle quality (Low-resolution reticle writer) | **> 6%** (Min. pillar D < 550 nm) | **~1** (Min. pillar D < 550 nm) |
| | Inner fields only | Inner reticles: Low-resolution Outer reticles: High-resolution | **> 44%** (Min. pillar D < 550 nm) | **> 0.67** (Min. pillar D < 550 nm) |
| | Outer fields only | Inner reticles: High-resolution Outer reticles: Low-resolution | **>39%** (Min. pillar D < 550 nm) | **> 0.64** (Min. pillar D < 550 nm) |
| **Pillar diameter shift** | All | Fabrication error | **> 68%** ($D \pm 50$ nm) | **~1** ($D \pm 50$ nm) |
| | Individual fields | Reticle quality control | **> 77%** ($D \pm 50$ nm) | **> 0.98** ($D \pm 50$ nm) |
| | Inner fields only | Inner reticles: Low-resolution Outer reticles: High-resolution | **> 74%** ($D \pm 50$ nm) | **> 0.96** ($D \pm 50$ nm) |
| | Outer fields only | Inner reticles: High-resolution Outer reticles: Low-resolution | **> 73%** ($D \pm 50$ nm) | **> 0.96** ($D \pm 50$ nm) |
| **Pillar height shift** | All | Fabrication error (error in etch rate) | **> 48%** ($H \pm 500$ nm) | **~1** ($H \pm 500$ nm) |
| **Non-uniform illumination** | All | Measurement setup error | N/A | **> 0.99** (Gaussian $w > 38$ mm) |

**Table. S3. Summary of simulation results on various fabrication and characterization error scenarios.**
High-resolution and low-resolution reticles correspond to industry-grade and in-house fabricated reticles, respectively. The main contributing factor to the aberrated focal spot obtained in experiment is the missing smaller pillars in the outer fields.

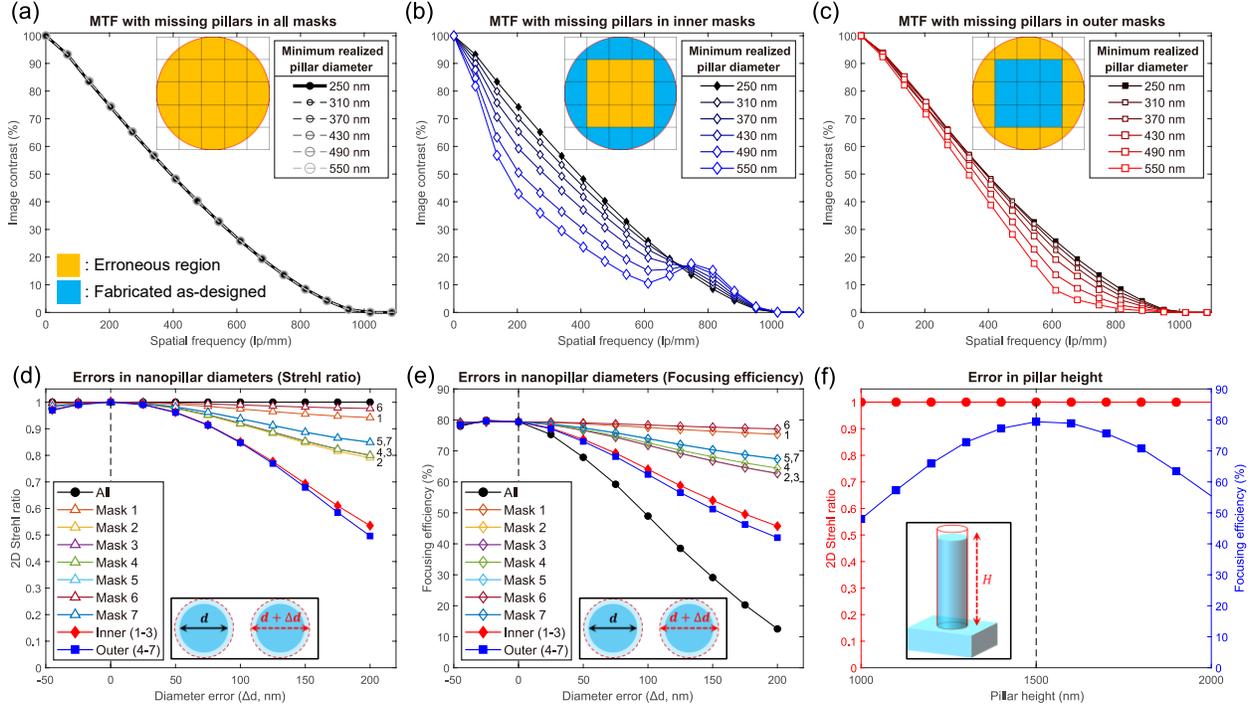

**Fig. S13. Effects of fabrication errors on the metalens imaging quality (simulation results).** MTF contrast curves with respect to minimum resolved nanopillar diameter in (a) all sections, (b) only the inner sections (reticles 1-3), and only the outer sections (reticles 4-7). All MTF curves are normalized to their zero-frequency values, respectively. The MTF degradation of (c) resembles that of the experimental data provided in the main text, which corroborates the SEM images and the diffracted intensity data. Effect of uniform offset in nanopillar diameters in all sections, individual sections, inner sections, and outer sections, respectively, on (d) Strehl ratio and (e) focusing efficiency, are shown, respectively. The metalens behaves diffraction limited (Strehl ratio > 0.8) even if the error in diameter is around 100 nm. This is due to nanopillars having a near-linear phase change with respect to their diameter (Fig. S2a), that each section experiences low phase aberration despite the overall diameter error. However, as the diameter offset between sections induce phase offsets between the sections (Fig. S2c), the focusing efficiency suffers due to out-of-phase interference at the focal plane between the sections. (f) shows effect of having overall error in nanopillar height on the Strehl ratio and focusing efficiency of the metalens. As the phase response of nanopillars with respect to height changes almost linearly (Fig. S2d), the Strehl ratio remains near unity even at with ±33% error or ±500 nm. However, when the height of the nanopillars deviates from the designed height of 1500 nm, the relative phase between the nanopillars and the empty area no longer meets the design, leading to a loss in focusing efficiency due to out-of-phase interference. We expect that if we can fill the metalens entirely with nanopillars, such loss of focusing efficiency by height error can be minimized.

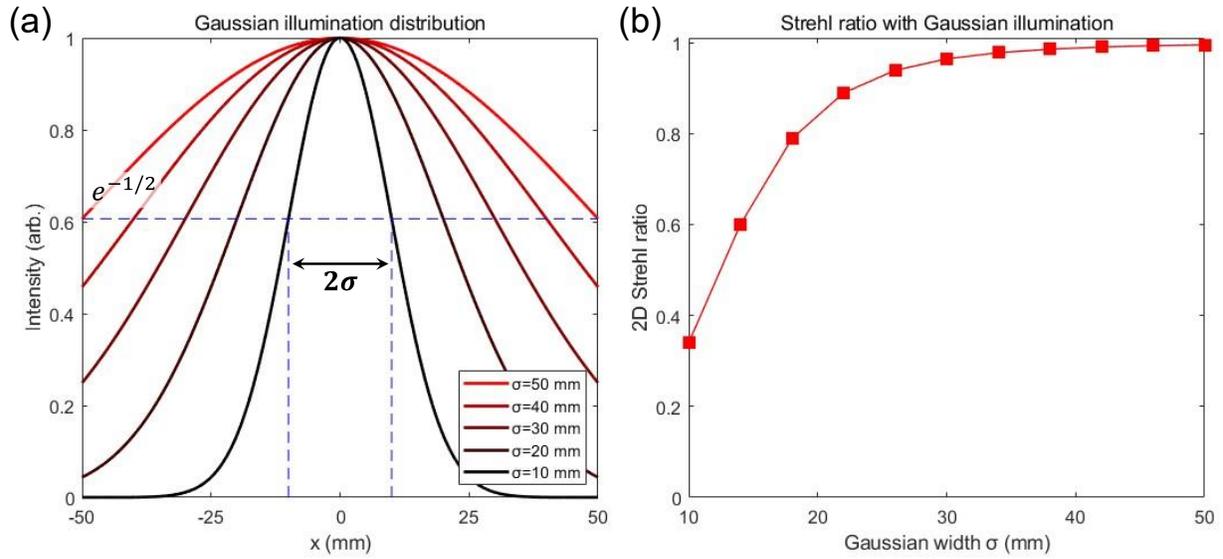

**Fig. S14. Effects of non-uniform illumination intensity on the metalens imaging quality (simulation results).**

We investigate the effect of non-uniform illumination intensity by modelling the incident intensity profile using a cylindrically symmetric Gaussian intensity profile with a flat phase front. The Gaussian width $w$ is defined so that the radial intensity profile has the form $I(\rho) = I_0 \exp(-\frac{\rho^2}{w^2})$. The Strehl ratio drops below the diffraction limit (0.8) when the Gaussian width of the incident beam is less than 40% of the total aperture size.

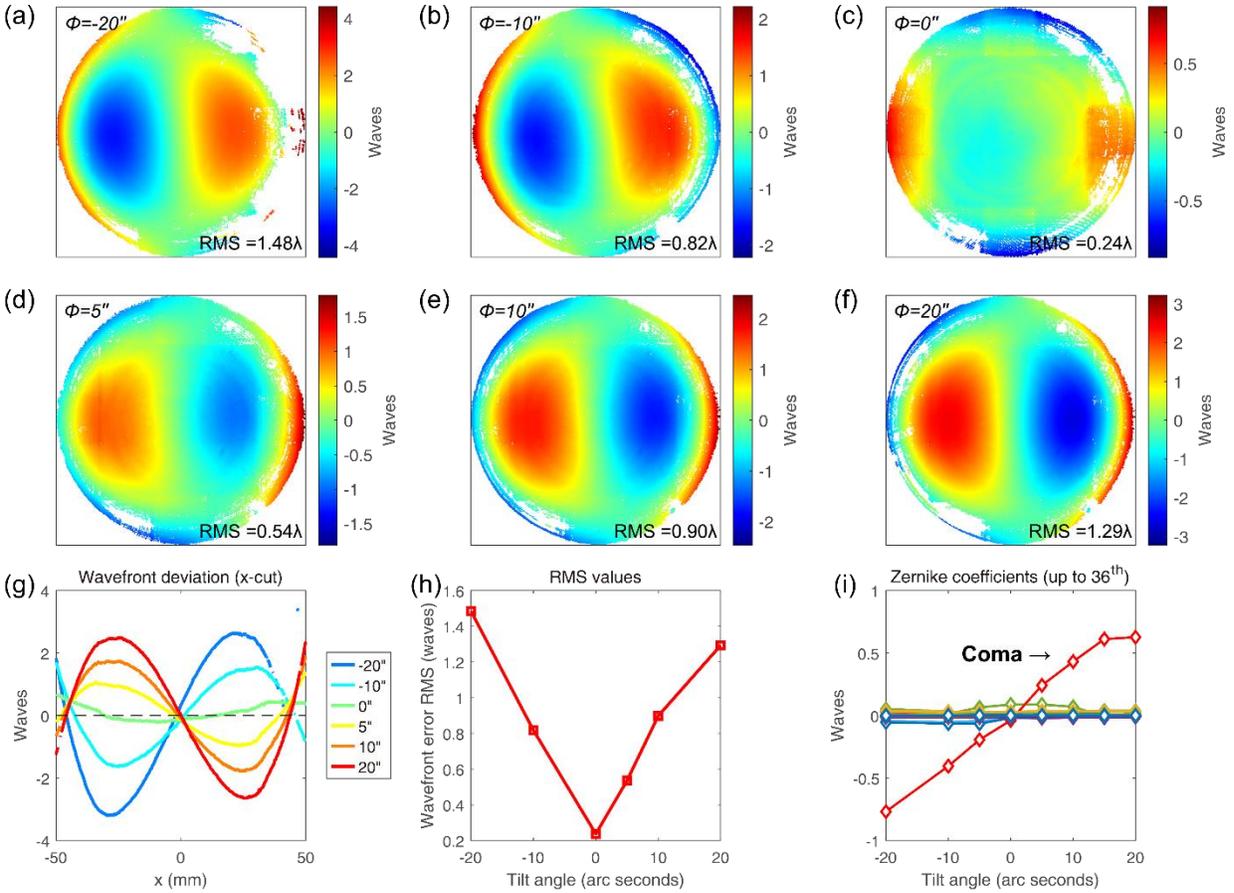

**Fig. S15. Measured wavefront error map with respect to various tilted incidence angles.**
Measured wavefront aberration from null interferometry configuration (Fig. S10) when metalens is tilted from normal incidence. This data represents and confirms the expected field-dependent aberration from a single optical component. Also, this measurement provides alignment tolerance in tilt when the metalens is placed in an imaging system. (a)-(f) Wavefront error of the metalens at various tilt angles ranging from -20" (arc seconds) to +20" after piston, tip, tilt, and power are removed. Due to low diffraction efficiencies at the outer region, interference information from certain points is not resolved. (g) Plots of wavefront error along the x-axis of the metalens at each measured incidence angles, respectively. (h) RMS of the wavefront error with respect to metalens tilt angles. (i) Plots of fitted Zernike polynomial coefficients up to 36th order with respect to the metalens tilt angles. Due to the lack of information at the edge of the metalens, fittings were performed using data from the central 80 mm diameter aperture area. The coma aberration is the dominant wavefront error with respect to lens tilt.

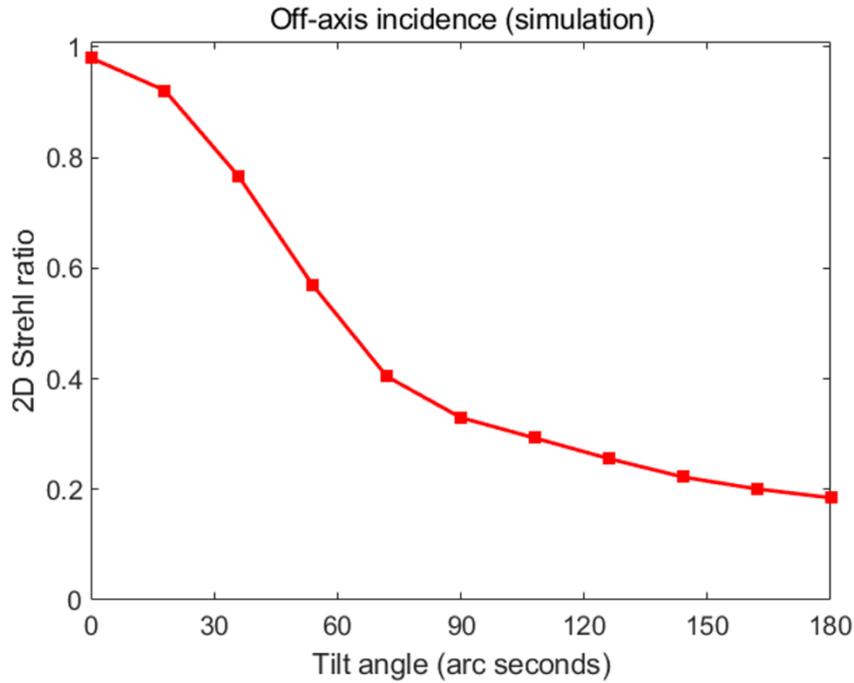

**Fig. S16. Simulated Strehl ratio of the 100 mm diameter metalens at various incident angles**. For this tilt study, the metalens focal spot is sampled on a $601 \times 601$ square grid around the tilted focal spot center up to a transverse diameter of 18 Airy disks. This large sampling region is required to capture the spatial patterns associated with strongly aberrated focal spots. The incident tilt angle is defined as the angle of the incident wavefront in air before it hits the first air/glass interface on the back face of the metalens. The Strehl ratio falls below the diffraction limit (~0.8) at a 30" tilt.

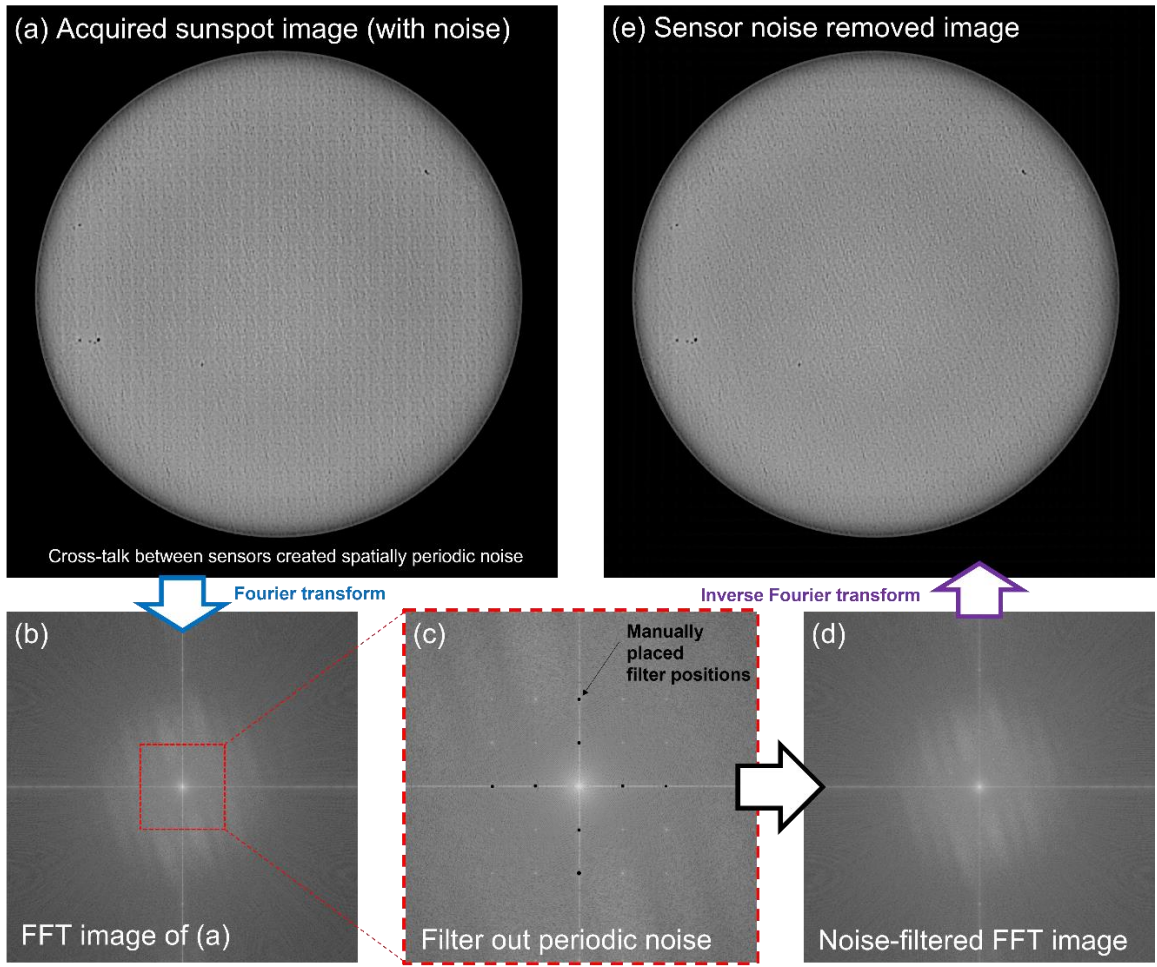

**Fig. S17. Sensor-induced noise removal with FFT for Sunspot imaging.**

(a) Strong intensity incidence from the Sun caused crosstalk between pixels of the camera sensors due to the on-sensor microlens array. Image taken on Feb. 1, 2023, 12:37 PM, on the rooftop of the Science Center building, Harvard University, Cambridge, Massachusetts. The meta-astroimager was mounted on a ZWO AM5 equatorial mount in sun tracking mode. Two filters, a 1 nm bandwidth $\lambda = 632.8\ nm$ bandpass filter (FL632.8-1, *Thorlabs*) and a neutral density filter of optical density (OD) 3.0, were placed in front of the image sensor. Imaging was performed using a ZWO ASI183mm-pro cooled CMOS imaging sensor (monochrome, 13.2 mm $\times$ 8.8 mm sensor, 2.4μm $\times$ 2.4μm pixel size) cooled to 10.5°C, with 8 ms exposure time per frame in RAW16 format and with a gain value of 85. Pixel binning was not used. A total of 1000 frames were used to process the image, using Astrosurface software (http://astrosurface.com/ U2 2023-01-25) with following parameters: Algo C quality estimator, 50 pixel tracking shift max, Planet/Disk target, 50% (500 frames) stack, stacking method by mean, global alignment mode, and sharpening using Wavelet-Deconvolution algorithm built in the Astrosurface software. (b) Fast Fourier transform (FFT) image obtained from (a) using ImageJ software (https://imagej.net). (c) Filter positions in *k*-space image to reduce low-order periodic noise. (d) FFT image after the noise filtering is applied. (e) Inverse-FFT image of (d), having some of the pixel noise removed. The sunspot image taken by public sources on the same date can be found online for comparison (https://soho.nascom.nasa.gov/data/synoptic/sunspots_earth/sunspots_512_20230201.jpg).

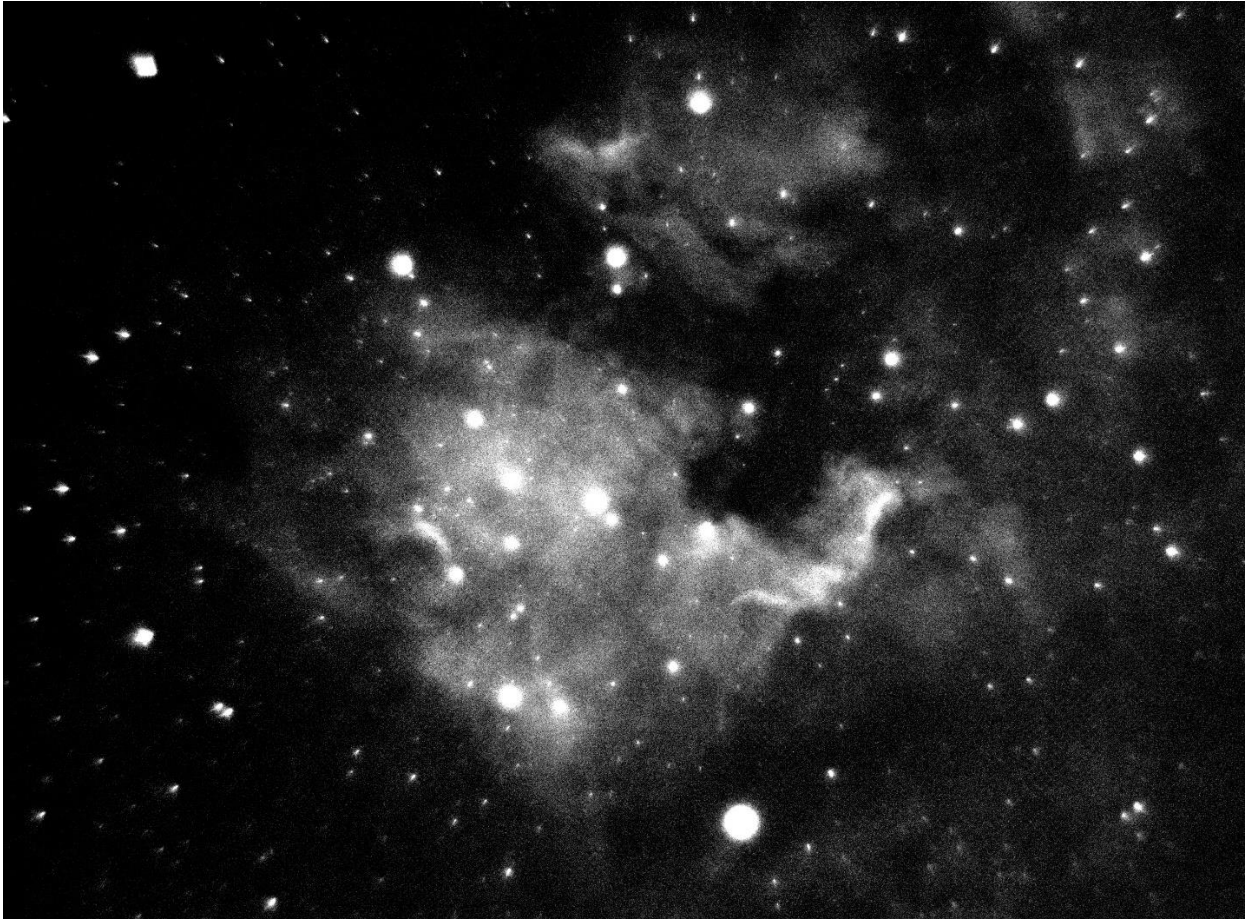

**Fig. S18. Acquired high-resolution image of the North America Nebula (NGC7000).**

Image of the North America Nebula (NGC7000) acquired from rooftop of the Science Center Building, Harvard University, Cambridge, Massachusetts, USA, on May 13, 2022. The used metalens astro-imager comprises a 100 mm diameter metalens, SVBONY 2-inch, 7 nm bandwidth H-alpha filter, and a ZWO ASI1600mm-pro cooled CMOS mono camera (4656 x 6520 pixels, 3.8 μm × 3.8 μm pixel size), mounted on a Celestron AVX equatorial mount with optical tracking provided from a QHY miniquidescope as a guide-scope mounted to a guide-camera (ASI178mm, ZWO). The image sensor was cooled to 0.5 °C, and 11 light frames and 4 dark frames were acquired with 120 second exposure time per frame in RAW12 format with a gain value of 200. Pixel binning was not used. The acquired images are then processed using Deep Sky Stacker (http://deepskystacker.free.fr/) software, with standard mode (Light: average, Dark: median, Alignment: Automatic).

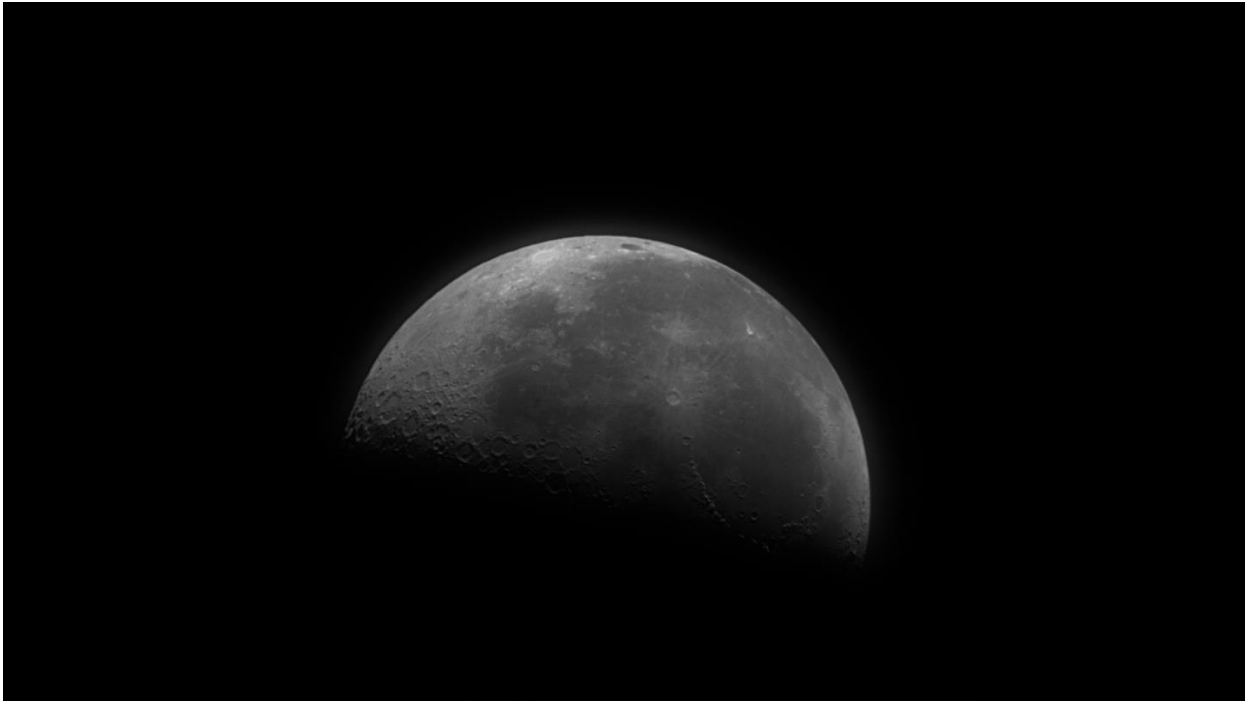

**Fig. S19. Acquired high-resolution image of the moon.**
Image of the Moon acquired from rooftop of the Science Center Building, Harvard University, Cambridge, Massachusetts, USA, on August 18, 2022. The used metalens astro-imager comprises a 100 mm diameter metalens, a 1 nm bandwidth bandpass filter at $\lambda = 632.8 \; nm$ (FL632.8-1, Thorlabs), and a ZWO ASI183mm-pro cooled CMOS mono camera (monochrome, 13.2 mm × 8.8 mm sensor, 2.4 μm × 2.4 μm pixel size) with sensor temperature at -0.5℃, mounted on an iOptron ZEQ25 equatorial mount controlled by ZWQ AsiAir Plus controller. Total of 413 frames were acquired with 442 ms exposure time per frame in RAW8 format and a gain value of 111. Pixel binning was not used. The acquired images are processed using Autostakkert software (https://www.autostakkert.com/, Image stabilization: Planet (COG), Dynamic background, Quality estimator: Local (AP), Frames to stack: 46% (189 frames), Drizzle 1.5x) and wavelet deconvolution was performed with Registax 6 software (https://www.astronomie.be/registax/).

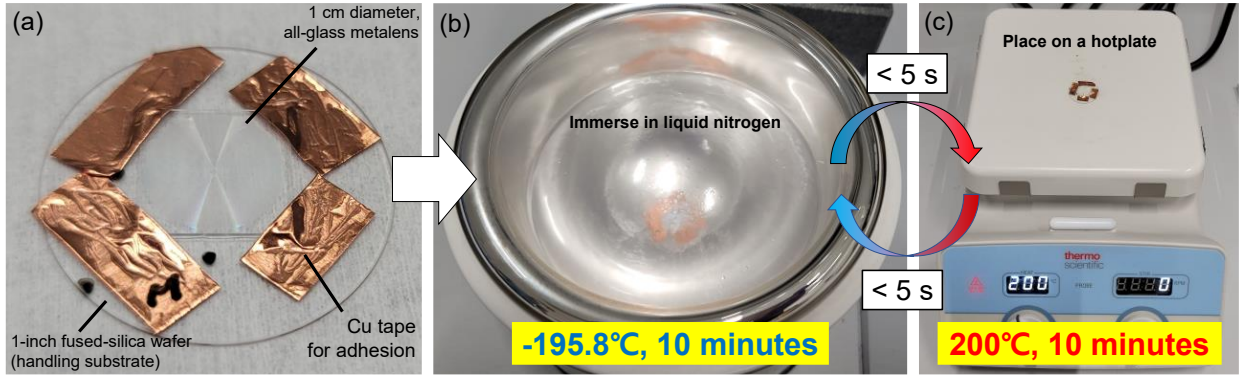

**Fig. S20. Extreme environment testing (thermal stress cycles) with 1 cm diameter all-glass metalens.**

A 1 cm diameter all-glass metalens (Ref. 1) was used as a proxy to test the glass nanopillar metalens' robustness against extreme temperature shocks that may occur when deployed in an extreme environment such as space. The shock test presented here approximates the United States military specifications (MIL-STD-883F METHOD 1011.9 and 1010.8), in which the sample goes through several cycles of thermal shock between hot and cold thermal reservoirs. (a) A 1 cm diameter metalens, diced into $1.1 \times 1.1$ cm square, is mounted on a 0.5 mm thick 1-inch diameter fused silica substrate (JGS2) for handling purposes. Copper (Cu) tapes were used to hold the metalens in place. (b) Metalens immersed in liquid nitrogen (-195.8℃) which serves as a cold reservoir. The sample remained in liquid nitrogen bath for 10 minutes to ensure reaching thermal equilibrium. (c) The cooled sample from (b) is then put on a hot plate (200℃) within 5 seconds after being taken out of the cold reservoir. The sample is kept on the hot plate for 10 minutes to ensure reaching thermal equilibrium. The sample is then put back in the liquid nitrogen bath within 5 seconds after it has been taken off the hot reservoir, where the process is repeated for total of 10 cycles, following the MIL-STD-883F METHOD 1011.9.

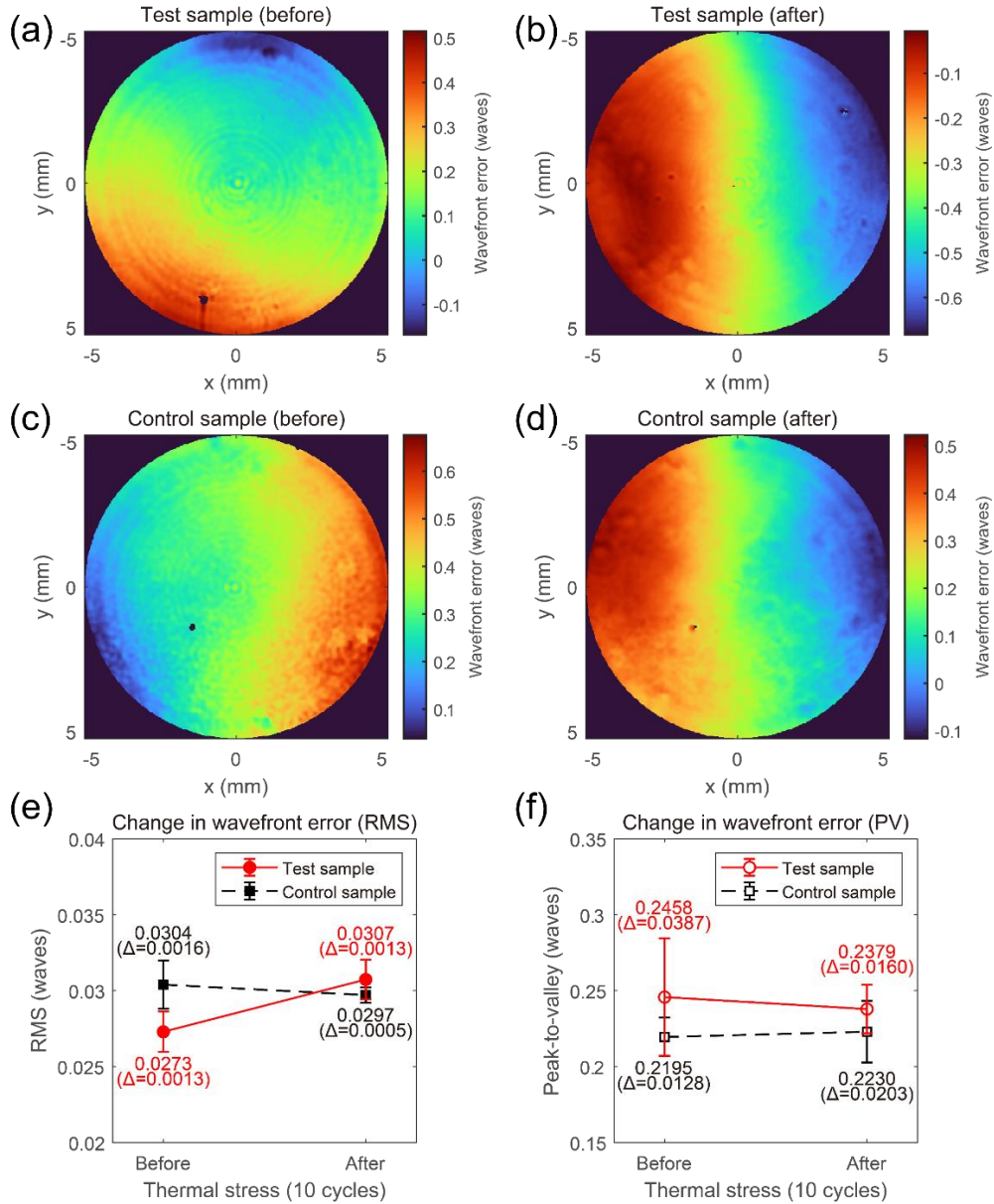

**Fig. S21. Result of extreme environment testing (thermal stress) with 1 cm diameter all-glass metalens.**

The measured wavefront error of the 1 cm diameter metalens before and after 10 cycles of thermal shock test in Fig. S20 are presented. Wavefront error of the metalens (a) before and (b) after thermal shock cycles. (c) and (d) show wavefront error of another 1 cm diameter metalens (control sample), which did not go through the thermal shock test. (e) Change in measured wavefront error's root-mean-squared (RMS) values before and after thermal shock, compared with that of the control sample. The plotted values are the mean and the standard deviation of 5 independent measurements for each data point, respectively. The RMS value appears to increase slightly, but it is within measurement error. (f) Measured peak-to-valley (PV) value of the wavefront error before and after thermal shock test compared with that of the control sample.

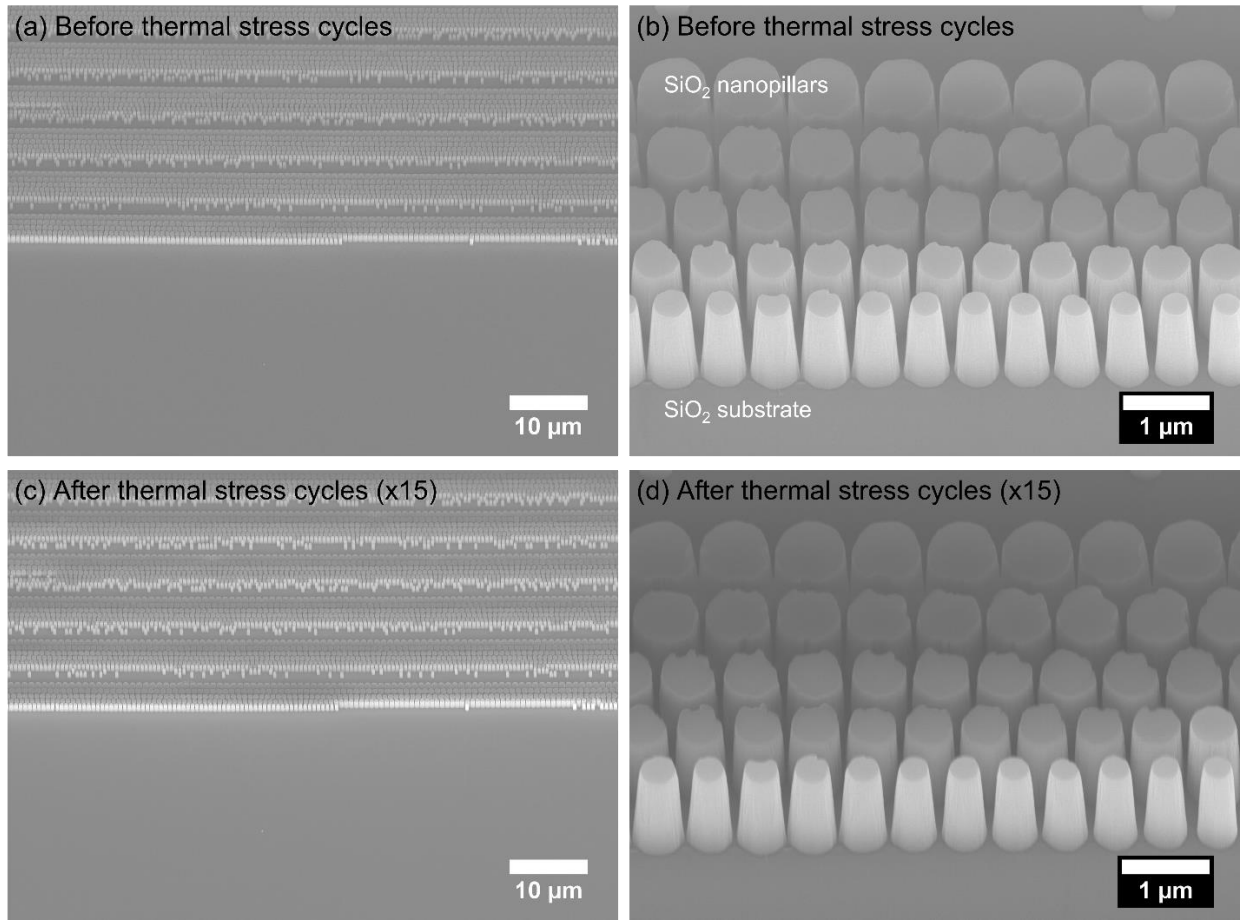

**Fig. S22. SEM images of 1 cm diameter all-glass metalens before and after extreme environment testing (thermal stress).**
(a) and (b) show the SEM images of the edge of the 1 cm diameter all-glass metalens, before going through the thermal stress cycles. (c) and (d) show the SEM images taken at the same region shown in (a) and (b). No apparent physical damage was observed even after 15 cycles of the thermal shock cycles. Gaps in the rings are where the nanopillars were not fabricated due to photomask resolution limitations. We do not observe evidence of fabricated nanopillars being damaged they leave a telltale mark at the base when they are broken off from the substrate.

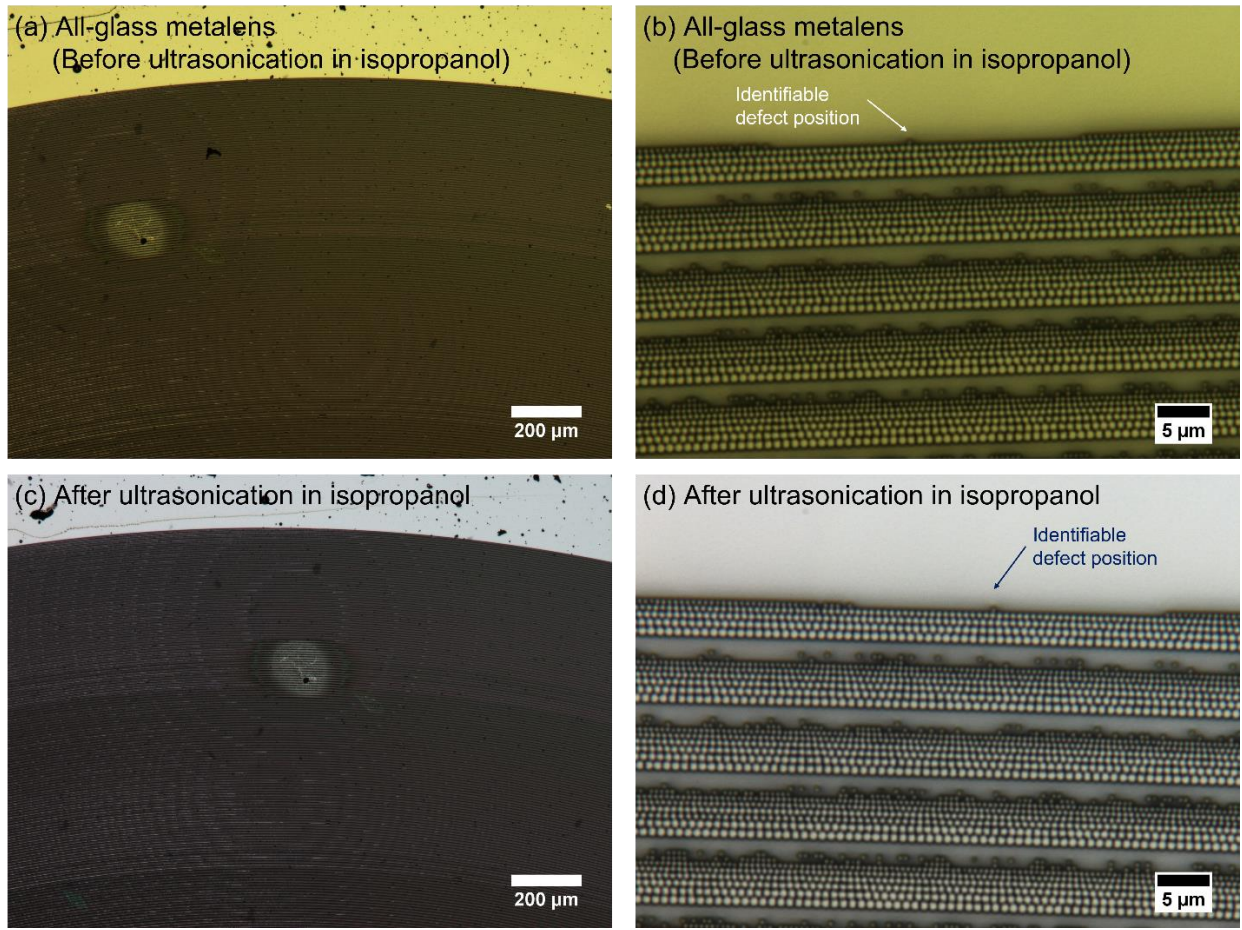

**Fig. S23. Optical microscope images of 1 cm diameter all-glass metalens before and after extreme environment testing (vibrational stress).**

A 1 cm diameter metalens was immersed in isopropanol and put in an ultrasonication bath (Ultra Clean Equipment, Inc.) for 20 minutes at the highest power setting to see effects of the vibrational stress on the nanopillars. From optical microscope images taken (a), (b) before and (c), (d) after ultrasonication, the nanopillars seem to remain intact even after prolonged exposure to sonic waves. The optical images are taken at a similar location to show that the nanopillars remain intact. Gaps in the rings are where the nanopillars were not fabricated due to photomask resolution limitations. We do not observe evidence of fabricated nanopillars being damaged they leave a telltale mark at the base when they are broken off from the substrate.

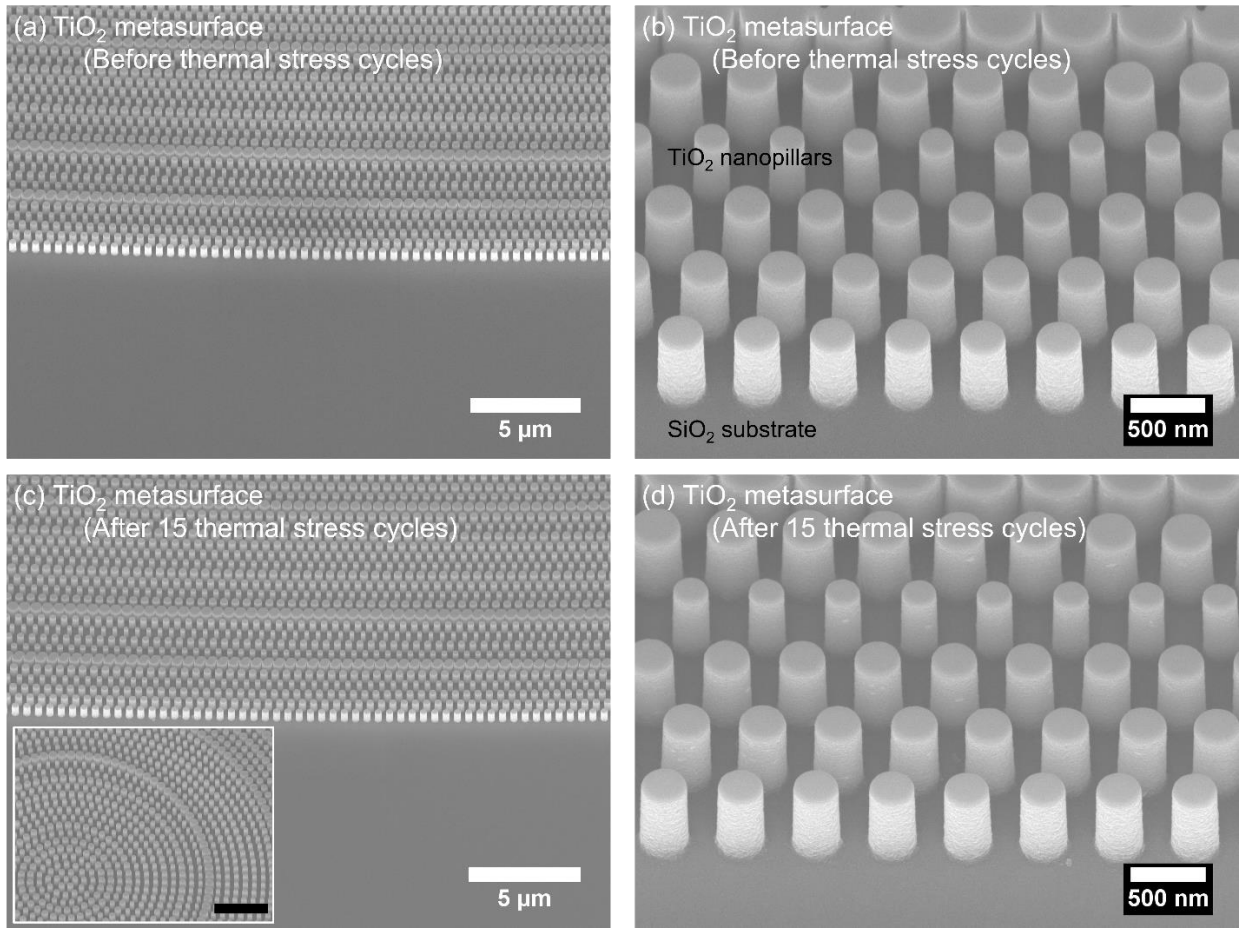

**Fig. S24. SEM images of 1 mm diameter TiO₂ metasurface before and after extreme environment testing (thermal stress).**

A 1 mm diameter TiO$_2$ metasurface (700 nm tall, Ref. 43) on a 1-in. diameter fused silica wafer went through the same thermal shock cycles, illustrated in Fig. S20, showing their robustness against rapid temperature fluctuations. The scalebar in the inset of (c) is 3 μm. More in-depth and controlled environmental testing will need to be performed to further confirm the robustness.

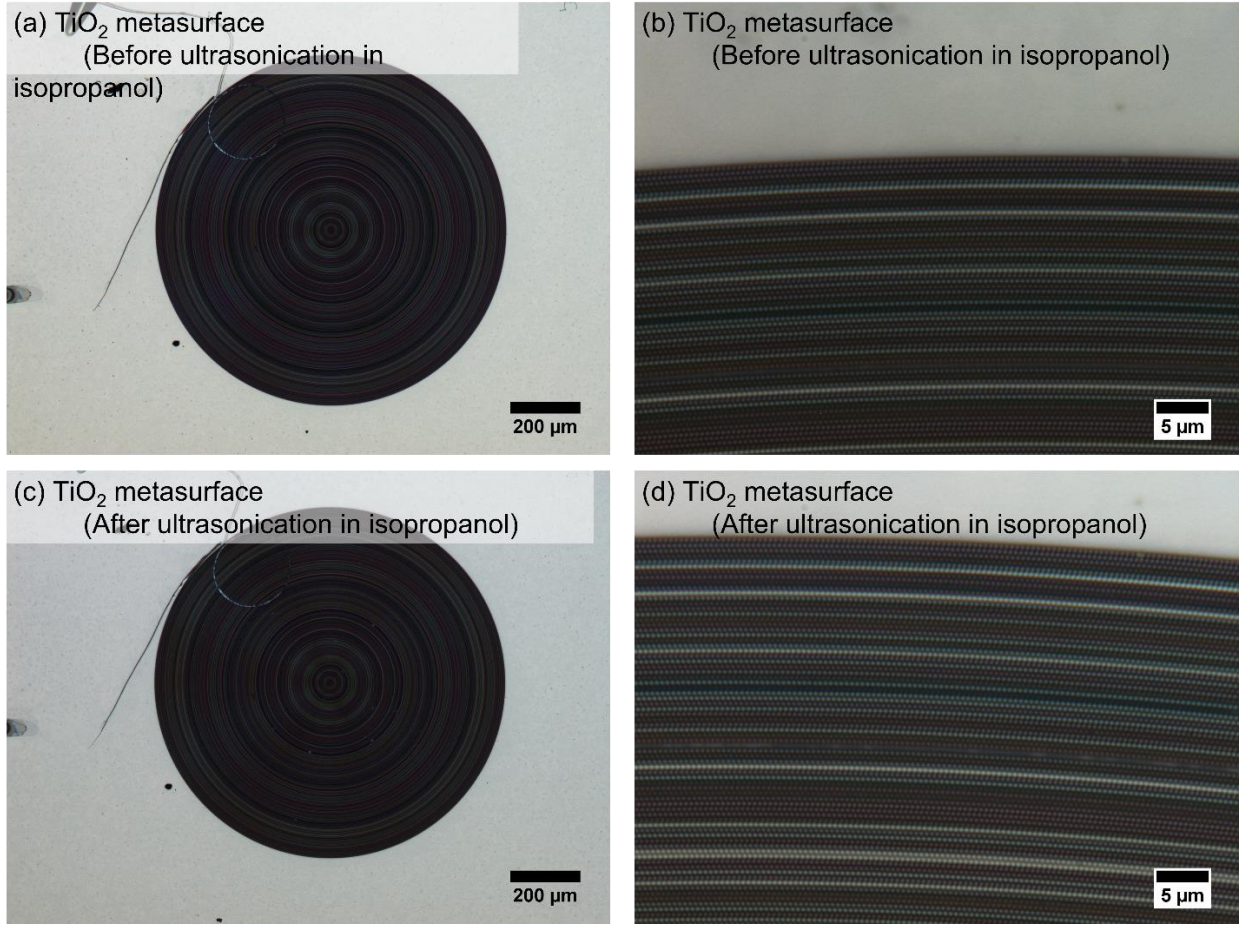

**Fig. S25. Optical microscope images of 1 mm diameter TiO₂ metasurface before and after extreme environment testing (vibrational stress).**

A 1 mm diameter TiO$_2$ metasurface (700 nm tall, Ref. 43) fabricated with e-beam lithography on a 1-in. diameter fused silica wafer went through the same 20 minute ultrasonication as illustrated in Fig. S23, which show their robustness against sonic waves. More in-depth and controlled environmental testing will need to be performed to further confirm their robustness.

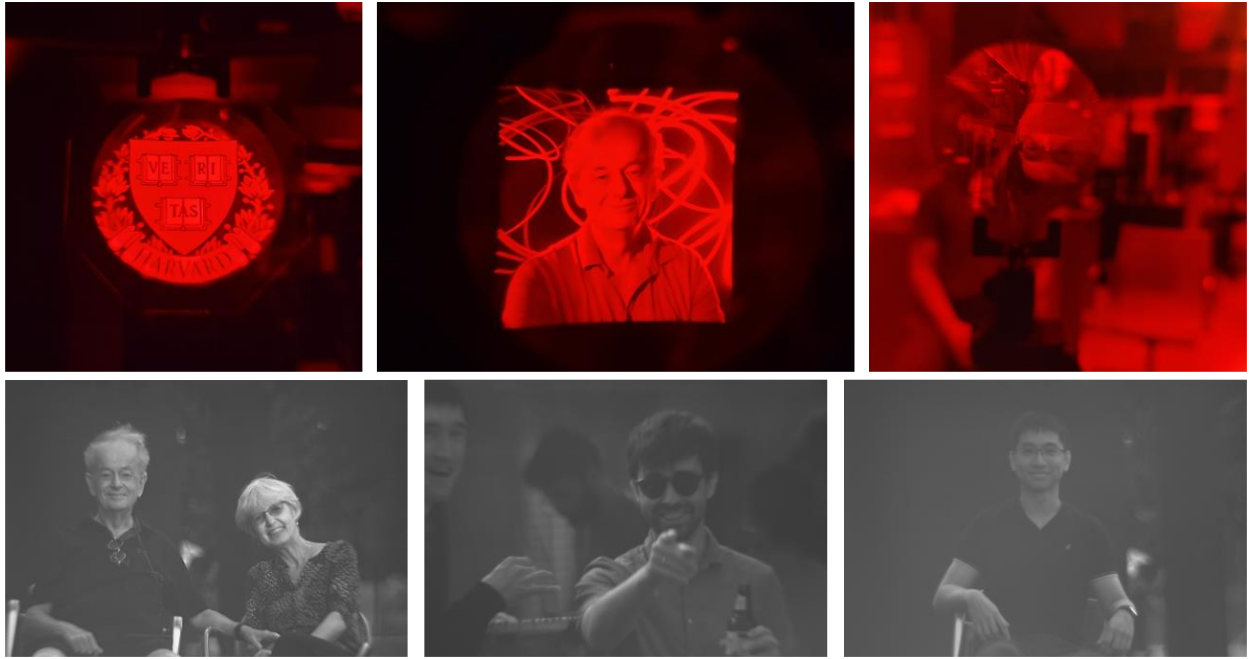

**Fig. S26. Other acquired images with the metalens (top row) and the meta-astrophotography apparatus (bottom row)**

(Top row) Images through the 100 mm metalens taken with a smartphone camera (Galaxy Note 20 Ultra, *Samsung*) with a 10 nm bandwidth 633 nm bandpass filter. Left two images are projected from a digital monitor (Galaxy Tab S7 Plus, *Samsung*). Image on the right shows an inverted image of the person (Soon Wei Daniel Lim) sitting on the other side of the lens.

(Bottom row) Images taken with the meta-astrophotography apparatus, with adjusted focus using the helical focuser. Left photo is Federico and Paola Capasso. Middle photo is Arman Amirzhan (left) and Vincent Ginis (right), and right photo is Soon Wei Daniel Lim. All images are taken by Joon-Suh Park.